\documentclass{article}
\usepackage{arxiv}

\usepackage[utf8]{inputenc} % allow utf-8 input
\usepackage[T1]{fontenc}    % use 8-bit T1 fonts
\usepackage{hyperref}       % hyperlinks
\usepackage{url}            % simple URL typesetting
\usepackage{booktabs}       % professional-quality tables
\usepackage{amsfonts}       % blackboard math symbols
\usepackage{nicefrac}       % compact symbols for 1/2, etc.
\usepackage{microtype}      % microtypography
\usepackage{lipsum}
\usepackage{graphicx}
\usepackage{enumitem}
\usepackage{gensymb}

\usepackage{multirow}%
\usepackage{amsmath,amssymb}%
\usepackage{amsthm}%
\usepackage{mathrsfs}%
\usepackage[title]{appendix}%
\usepackage{xcolor}%
\usepackage{textcomp}%
\usepackage{manyfoot}%
\usepackage{algpseudocode}%
\usepackage{listings}%

\usepackage{float}
\usepackage{tabularray}
\definecolor{customgray}{HTML}{dad7cd}
\definecolor{custompink}{HTML}{FB6F92}

\newcommand{\placeholdercite}[1]{\textbf{[REF]}}
\newcommand{\placeholderref}[1]{\textbf{[XXX]}}
\newcommand{\placeholder}[1]{\textbf{[PLACEHOLDER]}}

\DeclareRobustCommand{\rchi}{{\mathpalette\irchi\relax}}
\newcommand{\irchi}[2]{\raisebox{\depth}{$#1\chi$}}
\newcommand{\exponent}[1]{\ensuremath{\text{exp}\left(#1\right)}}

\usepackage{placeins}
\usepackage{mathtools}
\DeclarePairedDelimiter{\nint}\lfloor\rceil

\usepackage{physics}

\makeatletter
\newcommand*{\textlabel}[2]{%
  \edef\@currentlabel{#1}% Set target label
  \phantomsection% Correct hyper reference link
  #1\label{#2}% Print and store label
}
\makeatother

\usepackage[algo2e,ruled,linesnumbered]{algorithm2e}
\makeatletter
\newcommand{\nosemic}{\renewcommand{\@endalgocfline}{\relax}}% Drop semi-colon ;
\newcommand{\dosemic}{\renewcommand{\@endalgocfline}{\algocf@endline}}% Reinstate semi-colon ;
\newcommand{\pushline}{\Indp}% Indent
\newcommand{\popline}{\Indm\dosemic}% Undent
\let\oldnl\nl% Store \nl in \oldnl
\newcommand{\nonl}{\renewcommand{\nl}{\let\nl\oldnl}}% Remove line number for one line
\makeatother

\usepackage[backend=biber,style=numeric-comp,sorting=none]{biblatex}
\title{Explicit and fully automatic analysis of magnetotactic bacteria motion reveals the magnitude and length scaling of magnetic moments}

\author{
  Mara Smite\\
  MMML lab, Department of Physics\\
  University of Latvia (UL)\\
  Riga, Latvia, Jelgavas 3, 1004 \\
  \texttt{mara.smite@lu.lv} \\
  \And
  Mihails Birjukovs\\
  MMML lab, Department of Physics\\
  University of Latvia (UL)\\
  Riga, Latvia, Jelgavas 3, 1004 \\
  \texttt{mihails.birjukovs@lu.lv} \\
  \And
  Peteris Zvejnieks\\
  Department of Physics\\
  University of Latvia (UL)\\
  Riga, Latvia, Jelgavas 3, 1004 \\
  \texttt{peteris.zvejnieks@lu.lv} \\
  \And
  Ivars Drikis\\
  MMML lab, Department of Physics\\
  University of Latvia (UL)\\
  Riga, Latvia, Jelgavas 3, 1004 \\
  \texttt{ivars.drikis@lu.lv} \\
  \And
  Guntars Kitenbergs\\
  MMML lab, Department of Physics\\
  University of Latvia (UL)\\
  Riga, Latvia, Jelgavas 3, 1004 \\
  \texttt{guntars.kitenbergs@lu.lv} \\
  \And
  Andrejs Cebers\\
  MMML lab, Department of Physics\\
  University of Latvia (UL)\\
  Riga, Latvia, Jelgavas 3, 1004 \\
  \texttt{aceb@tesla.sal.lv}
}

\begin{document}
\maketitle

\begin{abstract}

Magnetotactic bacteria (MTB) are a diverse group of microorganisms whose movement can be directed via a magnetic field, which makes them attractive for applications in medicine and microfluidics. One of their key properties is the magnetic moment $m$, which is challenging to measure experimentally. We perform optical imaging experiments with MSR-1 MTB, and derive both the $m$ statistics and the scaling of $m$ with the MTB size using an explicit and fully automated method to determine $m$ from the MTB trajectories via the U-turn protocol, which measures $m$ based on the U-shaped trajectories exhibited by the MTB in an applied alternating magnetic field. The proposed method is an alternative to the standard U-turn time-based moment calculation and uses the theoretical U-turn shape function we have derived. This directly accounts for the U-turn geometry and determines the moment from the U-turn branch width. We couple this approach with a robust U-turn decomposition algorithm that detects U-turns from MTB tracks regardless of their orientations. We report a linear dependence of $m$ on the size of the bacteria, accounting for the bacteria velocity variations during the U-turns. We also demonstrate that the new U-turn shape-based and the conventional time-based methods produce significantly different results. The proposed method can be used to differentiate between various types of MTB within the same population based on their velocity and magnetic moments, and to precisely characterize the magnetic properties of a culture.

\end{abstract}


\clearpage

\section{Introduction}
\label{sec:introduction}

Magnetotactic microorganisms are a diverse group of single and multicellular organisms that are able to biomineralize magnetic nanoparticles, allowing them to passively align with an external magnetic field (MF), allowing their magnetic control \cite{mtb-review-paper-faivre}. They also propel themselves using a flagellar apparatus \cite{lefevre_diversity_2014,lauga-annurev-bacteria-hydrodynamics}. This group includes magnetotactic bacteria (MTB) \cite{Blakemore_1975,faivre_magnetotactic_2008}, multicellular magnetotactic prokaryotes \cite{lins_organization_1999, keim_swimming_2021}, as well as magnetotactic holobionts, a host organism that lives in symbiosis with magnetotactic bacteria \cite{sales_u-turn_2020,chevrier_collective_2023}. Magnetic microswimmers are of particular interest due to potential applications in drug delivery and cancer therapies, as MTB cells and magnetosomes are biocompatible and nontoxic \cite{mtb-drug-delivery, mtb-drug-delivery-more-recent, mtb-cancer-treatment-hyperthermia, fdez-gubieda_magnetotactic_2020, chen_bacteria-based_2023}. It is possible to form ensembles of swimmers and control their emergent behaviors, with a wide range of prospective applications in microrobotics \cite{lauga-annurev-bacteria-hydrodynamics, mtb-review-paper-faivre, msr-1-switching-swimming-modes, mtb-rotating-field-theory-erglis-cebers, mtb-rotating-field-synchorization-pairs, mtb-rotating-field-synchorization-swarms, cebers-ref-11, ClusterEmergence, mtb-tunable-self-assembly-swarms, mtb-tunable-hydrodynamics, mtb-confined-convection, mtb-aps-colloquium-review-faivre-2024, popp_polarity_2014}. For example, there is a clear potential for general-purpose object manipulation applications \cite{ClusterEmergence,mtb-tunable-hydrodynamics, mtb-tunable-self-assembly-swarms, cebers-ref-11}.

An important parameter of MTB and other magnetic microorganisms is their magnetic moment $m$, which can be determined by several methods, directly measuring the properties of the magnetic crystal or indirect measurements where $m$ is calculated based on how cells behave in the magnetic field \cite{nadkarni_comparison_2013}. For example, by observing the motion of the MTB in the rotating MF \cite{mtb-rotating-field-theory-erglis-cebers}, studying the fluctuations of the cell orientation \cite{smyk_orientation_fluct}, etc. Measurement of $m$ for a single cell can be performed by determining its velocity in an inhomogeneous MF. For example, by observing the motion of the MTB in a rotating MF, \textit{m} can be calculated from the fluctuations of the cell orientation  \cite{zahn_measurement_2017}. The most common approach is the \textit{U-turn method}, an indirect measurement proposed by Esquivel in 1985 \cite{Esquivel_motion_of_microorganisms}, which derives the magnetic moment from the trajectory of microorganisms in a pulsed MF \cite{chevrier_collective_2023,alvaros_u_turn_cocci,codutti_interplay_2022}. Bacteria are oriented and swim along the MF direction, and change the swimming direction upon MF reversal, creating the distinctive U-shaped trajectories. By determining the time $\tau$ it takes the microorganism to complete a U-turn, $m$ can be estimated as follows \cite{Esquivel_motion_of_microorganisms}:

\begin{equation}
\tau=\frac{8\pi\eta R^3}{mB} \cdot \ln{\left( \frac{2mB}{kT} \right)}
\label{eq:magnetic-moment-tau}\
\end{equation}
where $\eta$ is the environment (sample) viscosity, $R$ is the (effective) radius of a microorganism (assuming spherical shape), $B$ is the applied MF induction, $k$ is the Boltzmann constant and $T$ is the sample temperature. The above equation can be interpreted as the dependence of the U-turn time on the ratio of torques due to rotational drag (a spherical body is assumed) and realignment with reversed MF, with a correction term due to thermal fluctuations.

However, the issue is that the underlying derivation is based on physical assumptions that are not sufficiently general. Typically, $\tau$ is measured based on a fraction of a U-turn, which is arbitrary selected and can be subjected to considerable noise and the usually observed cell movement deviation from the MF direction. That is, this method does not explicitly determine $m$ from the shape of the U-turn, which would allow obtaining a more accurate estimate for $m$ based on the entire U-turn. Measurements of $\tau$ are most commonly performed semi-manually (e.g., manual selection of eligible trajectories and U-turns, and U-turn diameter estimates without shape fitting) \cite{Pichel_2018} or, as shown in \cite{alvaros_u_turn_cocci}, by detecting U-turn boundaries from extrema in U-turn trajectory coordinate derivatives
\cite{nadkarni_comparison_2013,sales_u-turn_2020,chevrier_collective_2023}. However, derivatives in only one direction are computed (vertical/horizontal), which is prone to low U-turn endpoint contrast in the coordinate time series, errors due to data noise, is generally problematic in the case of trajectories with an arbitrary configuration of U-turns, and does not account for U-turns tilted at an angle to the horizontal/vertical axis of the image and/or at an angle to applied MF. The latter is important because significant misalignment between the MTB magnetic moment and its body axis has been reported for \textit{Magnetospirillum magneticum} (AMB-1) \cite{nagard-mtb-body-moment-misalignment}, but also because the MTB trajectories may be perturbed by other means. In some cases, trajectory and U-turn analysis is not sufficiently documented \cite{david_soto_rodriguez_azide_2023}, making reproducibility an issue.

\begin{figure}[h!]
\begin{center}
\includegraphics[width=1.0\textwidth]{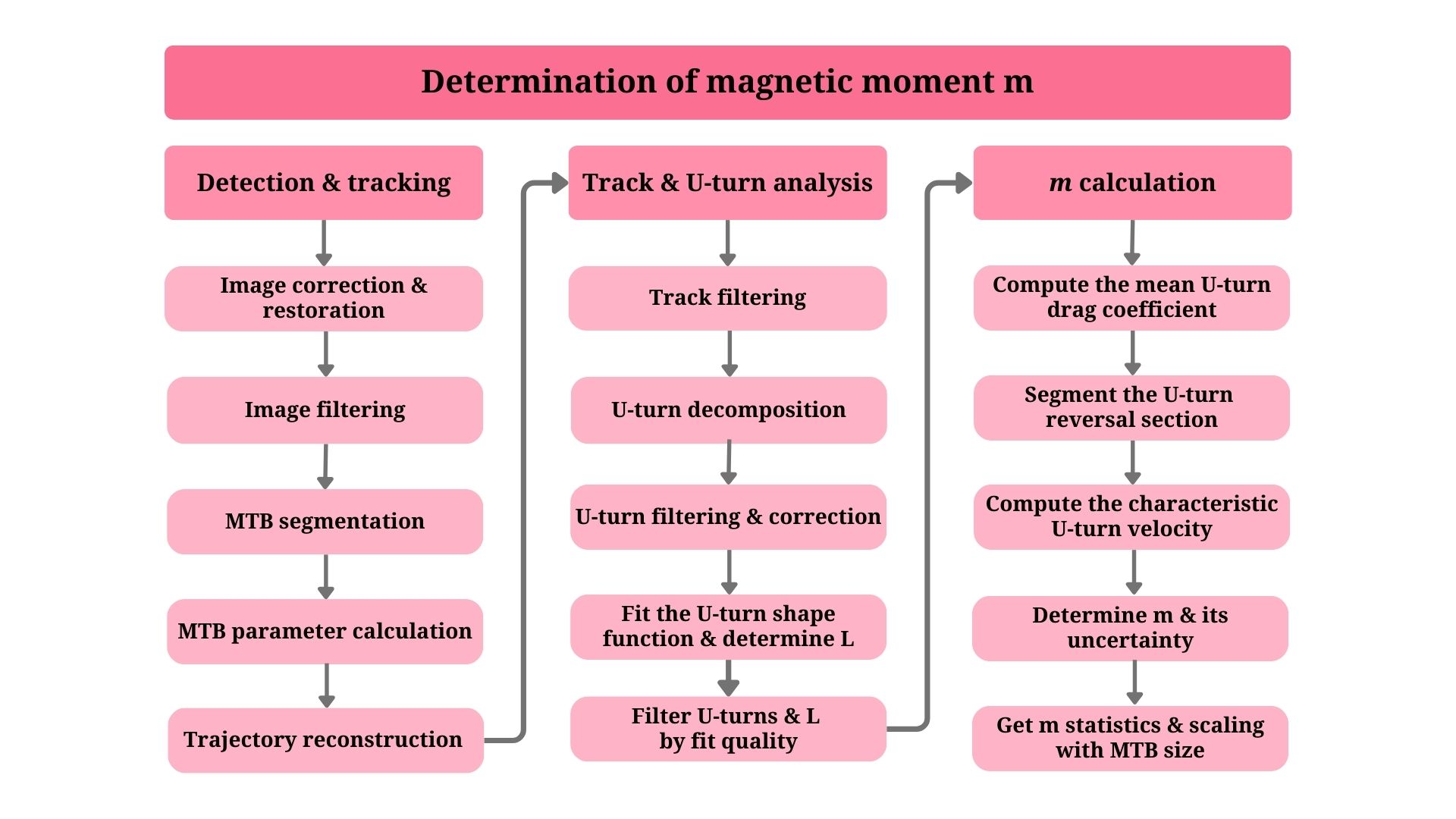}
\caption{An overview of the algorithm used to determine $m$ from MTB trajectories.}
\label{fig:algorithm-from-image-to-mtb-moments}
\end{center}
\end{figure}

We propose an explicit and fully automated method to determine $m$ from the trajectories of microorganisms with U-turns in alternating MF. We present a theoretical shape function for MTB U-turns in Section \ref{sec:theory-mtb-motion-model}, which enables computing $m$ explicitly from the width of the U-turn, circumventing $\tau$ calculations, as well as an algorithm for decomposing MTB trajectories into U-turns and determining $m$ for each instance (Section \ref{sec:determining-L} and Algorithm \ref{alg:mtb-moment-extraction}). We demonstrate the effectiveness of our approach on optically imaged MSR-1 U-turn motion in an alternating MF. We detect MTB from brightfield images, recover their trajectories, and apply the new approach to determine $m$ statistics and $m$ scaling with the MTB size, then compare the results with magnetic moment calculated using $\tau$. We also show the effects of different estimates for characteristic U-turn velocity values on the resulting $m$ statistics. The proposed data analysis pipeline is shown in Figure \ref{fig:algorithm-from-image-to-mtb-moments}.

\section{Materials and methods}
\label{sec:methods-materials}

\subsection{Cell cultivation}
\label{sec:cell-cultivation}

Flask standard medium (FSM) established by Heyen and Schüler \cite{heyen_growth_2003} was used to cultivate \textit{Magnetospirillum }\textit{gryphiswaldense} (MSR-1) magnetotactic bacteria. 100 $\mu l$ of cell suspension was inoculated into 16 $ml$ plastic centrifuge tubes filled with 15 $ml$ of FSM, ensuring microaerophilic conditions. Cells were incubated at $29^{\circ} C$ for 3 days. Bacterium suspension added directly from the growth tube was used in microscopy.

\subsection{Optical microscopy \& image acquisition}
\label{sec:microscopy-image-acquisition}

Microscopy samples were prepared by filling a glass capillary ($0.2~mm \times 2.0~ mm \times 16 ~mm$)with a suspension of bacteria. Both ends of the capillary were sealed with thermopaste, preventing the formation of an oxygen gradient and the formation of a microaerobic bacteria band. The total experiment time, starting with the removal of the culture from the incubator, was 90 minutes.

\begin{figure}[h]
\begin{center}
\includegraphics[width=0.5\textwidth]{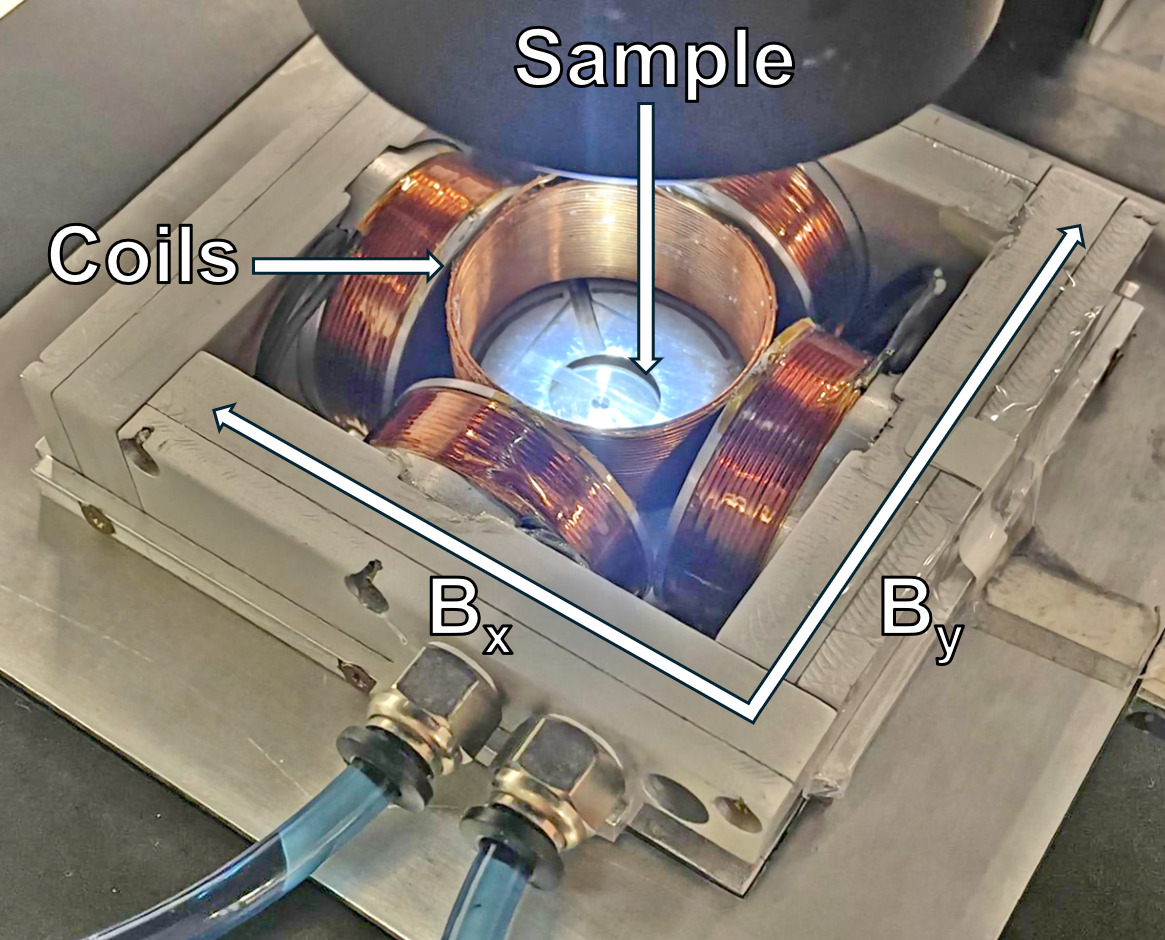} 
\caption{The microscopy setup: a microscope stage equipped with 3 coil pairs generating homogenous MF.}
\label{fig:experimental-setup}
\end{center}
\end{figure}

The capillary was placed on an optical microscope stage (\textit{Leica DMI3000B}) equipped with a $40 \times$ magnification objective. The microscope stage has three built-in coil pairs, generating a homogeneous MF in the sample (Figure \ref{fig:experimental-setup}). The MF is generated using an AC power supply (\textit{Kepco BOP 2010M}), and controlled via a data acquisition card and a \textit{LabVIEW} program. A \textit{Basler AC1920155UM} camera was used to record the cell motion.

\begin{figure}[h!]
\begin{center}
\includegraphics[width=0.4\textwidth]{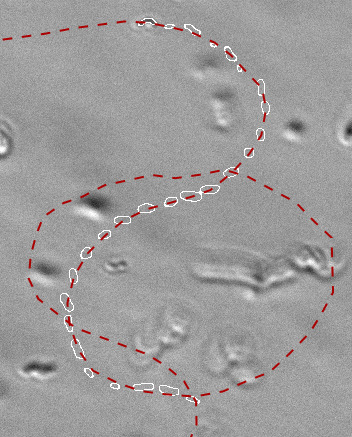} 
\caption{\textbf An MSR-1 bacterium trajectory in an alternating MF. The dashed line indicates the trajectory, and the white outlines show the cell shape as it is tracked for $\sim 1.35$ seconds.}
\label{fig:U-turns-example}
\end{center}
\end{figure}

The sample was placed in an alternating MF ($B=2.0~mT$, switching frequency $f=0.7~Hz$), causing the most motile and magnetic cells to change their swimming direction when the MF direction is reversed, thus inducing the characteristic U-turn motion (Figure \ref{fig:U-turns-example}). The image acquisition frame rate is set to $20$ frames per second (FPS). The image recording is triggered by the MF direction change. The duration of every constant MF direction interval is equal. The MF strength is calculated from the coil current measurements at the moment when camera exposure for each frame starts.

\subsection{MTB detection \& tracking}
\label{sec:detection-tracking}

To remove static background artifacts and features from the raw images, flat-field correction (FFC) is applied to images. This FFC procedure is adaptive and detects intervals of constant field of view (FOV) and image focus, splits the image sequence into these intervals, and then performs FFC with mean interval images for each sequence interval. This is because the FOV and focus are sometimes adjusted for a better view with more active and MF-compliant MTB, and to assess MTB behavior within different fluid layers. Constant FOV or focus intervals are detected as intervals between extrema in the normalized second derivative magnitude of the mean squared error (MSE) time series for consecutive pairs in the image sequence. To avoid overfitting insignificant background variations, the MSE time series is filtered with the Gaussian total variation (TV) filter \cite{total-variation-rof-model}, and the extrema are thresholded by relative magnitude. Other static artifacts and stuck MTB are removed by detecting them from the temporal standard deviation projection image computed from the corrected image sequence. This is done by applying color tone mapping (CTM) \cite{reproduction-of-color-chapter-6}, normalizing the image, binarizing with a manual threshold or using the Kapur's method \cite{kapur-entropy-segmentation}, slightly inflating the artifact masks via morphological dilation (circular kernel) \cite{images-mathematical-morphology}, and afterward the artifacts are removed via texture synthesis inpainting \cite{wolfram-mathematica-inpaint}. Then, to improve the contrast-to-noise ratio (CNR), local adaptive CTM and referenceless FFC are applied. Next, the Poisson model TV filter \cite{total-variation-poisson} is used for denoising. The Poisson TV filter is chosen because it effectively counteracts the additional luminance-dependent noise introduced during the above image correction stages. To segment MTB from filtered images, the CNR is further boosted by applying soft CTM masking (SCTMM) \cite{birjukovs2021resolving} (used for the detection of MTB swarms and individual MTB in \cite{mtb-swarms-theory-arxiv-2024}) and the maximum filter, and then Otsu binarization \cite{otsu-thresholding} is performed, followed by size thresholding to remove fine artifacts. Finally, any remaining segments of the ringing artifacts due to optical imaging are removed using the luminance-based false positive elimination method developed in \cite{birjukovs2021resolving, birjukovs-particle-EXIF} and previously applied to the detection of MTB in \cite{mtb-swarms-theory-arxiv-2024}. MTB segment coordinates and shape parameters are then determined, and the trajectory reconstruction is performed using the MHT-X object tracking code \cite{mht-x-og, birjukovs-particle-EXIF, birjukovs-particle-track-curvature-stats}. The resulting trajectories are then analyzed as outlined in Section \ref{sec:determining-L}. The parameters for the image processing and trajectory tracking codes are provided with the code, which is linked in the Code Availability section.

\subsection{MTB U-turn trajectory model}
\label{sec:theory-mtb-motion-model}

Let the orientation of MTB and its magnetic moment be $\vec{n}=( \cos{(\vartheta')},\sin{(\vartheta')},0)$, and the MF strength before/after reversal be $\vec{H}=(0,H,0)$ and $\vec{H}=(0,-H,0)$, respectively. Since in our experiment $H \ll 2\pi M$, we assume that the magnetic moment of the MTB is along its body axis \cite{mtb-rotating-field-theory-erglis-cebers}. An MTB moves in the $x,y$ plane, and its thermal fluctuations are negligible. The angular velocity of the MTB is $\vec{\Omega}=[\vec{n}\times\dot{\vec{n}}]=\dot{\vartheta'}\vec{e}_z$. 
The balance of the hydrodynamic and magnetic torques reads as follows:

\begin{equation}
-\alpha\vec{\Omega}+[\vec{m}\times\vec{H}]=0
\end{equation}
where $\alpha$ is the rotational drag coefficient and $\vec{m}=m\vec{n}$ is the MTB magnetic moment. 
Projected onto the $z$ axis, this becomes $\alpha\dot{\vartheta'}=-mH\cos{(\vartheta')}$. 
Introducing $\vartheta=\vartheta'-\pi/2$, the angle between the MTB and MF before its reversal, one has

\begin{equation}
\alpha\dot{\vartheta}=mH\sin{(\vartheta)}
\label{Eq:1}
\end{equation}

The velocity of the MTB translational motion is $\vec{v}=v\vec{n}$ and a constant $v$ is assumed. Introducing the trajectory arc length $l$, velocity $v = dl/dt$ and \eqref{Eq:1} may be rewritten as follows:

\begin{equation}
\frac{d\vartheta}{dl}=\frac{mH}{\alpha v} \cdot \sin{(\vartheta)}
\label{Eq:2}
\end{equation}

The asymptotic U-turn conditions are

\begin{equation}
    \lim_{l\rightarrow -\infty} \vartheta (l) = 0;~ \lim_{l\rightarrow + \infty} \vartheta (l) = \pi
\end{equation}
at the U-turn start and the U-turn end, respectively. Then, integrating \eqref{Eq:2}, the following is obtained \cite{zahn_measurement_2017}:
\begin{equation}
\tan{(\vartheta/2)}=\exponent{\frac{mHl}{\alpha v}}
\label{Eq:3}
\end{equation}

Solution \eqref{Eq:3} allows calculating the trajectory by integrating the equations of motion
\begin{equation}
\frac{dx}{dl}=-\sin{(\vartheta)};~\frac{dy}{dl}=\cos{(\vartheta)}
\label {Eq:4}
\end{equation}
using
$$
\sin{(\vartheta)}=\frac{2\tan{(\vartheta/2)}}{1+\tan^{2}{(\vartheta/2)}}
$$
and obtaining
\begin{equation}
\frac{dx}{dl}=-\frac{1}{\cosh{(\beta l)}}
\label{Eq:5}
\end{equation}
where $\beta = mH/\alpha v$. Integrating \eqref{Eq:5} yields 
\begin{equation}
x(l)=-\frac{2}{\beta} \cdot \arctan{\left( \tanh{(\beta l/2) } \right)} + C_1
\label{Eq:6}
\end{equation}
where $x(0)=0 \implies C_1=0$ and
\begin{equation}
x=-\frac{2}{\beta} \cdot \arctan{\left( \tanh{(\beta l/2) } \right)}
\label{Eq:7}
\end{equation}

Similarly, using
$$
\cos{(\vartheta)}=\frac{1-\tan^{2}{(\vartheta/2)}}{1+\tan^{2}{(\vartheta/2)}}
$$
an equation for the $y$ coordinate can be written as
\begin{equation}
\frac{dy}{dl}=-\tanh{(\beta l)}
\label{Eq:8}
\end{equation}

Integration over $l$ results in
\begin{equation}
y(l)=- \frac{1}{\beta} \cdot \log{\left( \cosh{(\beta l)} \right)} + C_2
\label{Eq:9}
\end{equation}
and, since $y(0)=0 \implies C_2=0$. Using
$$
\cosh{(\beta l)}=\frac{1+\tanh^{2}{(\beta l/2)}}{1-\tanh^{2}{(\beta l/2)}}
$$
and combining \eqref{Eq:7} with \eqref{Eq:9}, the trajectory equation is 
\begin{equation}
y=-\frac{1}{\beta} \cdot \ln{\left(\frac{1+\tan^{2}{(\beta x/2)}}{1-\tan^{2}{(\beta x/2)}}\right)}
\label{Eq:10}
\end{equation}

Since 

\begin{equation}
    \lim_{\beta x/2 \rightarrow \pm \pi/4} y(x) = -\infty
\end{equation}
the trajectory can be parametrized by introducing the asymptotic distance $L$ between the branches of the U-turn trajectory $L=\pi / \beta$. Then 
\begin{equation}
y=-\frac{L}{\pi}\ln{\left(\frac{1+\tan^{2}{(\frac{\pi x}{2L}})}{1-\tan^{2}{(\frac{\pi x}{2L})}}\right)}
\label{Eq:11}
\end{equation}
with 

\begin{equation}
    \lim_{x\rightarrow \pm L/2} y(x) = -\infty
\end{equation}
as it should be. Finally, we can simplify the above expression using the following expressions:

$$
2 \sin^2{\left( \frac{u}{2} \right)} = 1 - \cos{u};~~ 2 \cos^2{\left( \frac{u}{2} \right)} = 1 + \cos{u}
$$
to a more concise form

\begin{equation}
\boxed{
    y(x,L) = - \frac{L}{\pi} \ln{ \left( \sec \left( \frac{\pi x}{L} \right) \right) }
}
\label{eq:uturn-shape-equation-final}
\end{equation}

Given $L$ determined from an experimental trajectory, the MTB magnetic moment can be computed via

\begin{equation}
\boxed{
m=\frac{\pi\alpha v}{HL}
}
\label{eq:moment-from-uturn-params}
\end{equation}
if the rotational drag coefficient $\alpha$ and velocity of MTB $v$ over the U-turn are known. In our case, we assume the MTB are spherical and use the Stokes rotational drag, but in general any drag model can be used here -- for example, elongated bacteria can be approximated as a chain of spheres, for the which the rotational drag coefficient has been derived \cite{mtb-rotational-drag-sphere-chain-model}. Alternatively, MTB can be modelled as cylinders \cite{Pichel_2018} or ellipsoids \cite{leao_eliptic_drag_c}, or an experimentally determined spiral/spirillum drag coefficient can be used \cite{Pichel_2018}.

\subsection{Determining MTB magnetic moment from experimental data}
\label{sec:determining-L}

Generally, MTB tracks can have arbitrary shape despite consistent alternating MF, and may have different orientations due to magnetic moment and cell body misalignment \cite{nagard-mtb-body-moment-misalignment}, moment misalignment in magnetosome chains \cite{mtb-msr1-magnetosome-moment-misalignment-in-chains}, etc.), with many U-turns per track (Figure \ref{fig:U-turns-example}). Therefore, we must decompose the tracks into U-turns and fit the track points to \eqref{eq:uturn-shape-equation-final}, then determine $L$, $v$ and $\alpha$ for \eqref{eq:moment-from-uturn-params} from the resulting U-turns.

Before performing U-turn decomposition, tracks are filtered by temporal length to exclude undersampled MTB motion, then by minimum-oriented bounding box aspect ratio (oriented aspect ratio, excludes trajectories of MF that do not respond to MF changes, since they have very high aspect ratio values), and then filtered using median and mean filters (1 point-wide kernels).

The eligible filtered tracks are decomposed into U-turns by finding peaks in absolute values of the second derivative of their coordinate time series and thresholding the peaks by relative values, and extrema of curvature (their \textit{turning points}) within the Gaussian-filtered track (thresholded by the relative curvature value). Trajectories are then split into fragments at the coordinates of these points, resolving overlaps/proximity cases, if any. U-turns are then found among the resulting track fragments as those that contain the curvature extrema points, and the branches of the U-turns (points on either side of the turning points) are pruned to remove artifacts from the tip of the branch of the U-turn.

Among the U-turns generated from the trajectories, the eligible U-turns are found through a sequence of filtering procedures. After filtering U-turns by temporal length, instances with self-intersections are eliminated by converting the sequence of U-turn point into a graphics line object, then to a raster image with boundary padding; the image is then binarized via the Otsu method, Gaussian-filtered and binarized again to ensure that the rasterized line is unbroken, and afterward the number of domains in the image is counted -- for a U-turn without self-intersection, there is only a single domain (albeit not simply connected), whereas self-intersection instances generate additional domains. After removing self-intersecting U-turns, the remainder are filtered by minimum bounding box aspect ratio and area (this removes instances with stuck MTB and straighter fragments of tracks), and then the turning point locations are refined (turning points are detected for uniformly upsampled versions of U-turns, B-spline order 2), and U-turns with multiple turning points post-refinement are eliminated.

The surviving U-turns are symmetrized by physical length with respect to their turning points -- this is to minimize bias during theoretical shape fitting. Then the U-turns are translated such that their turning points are at $(x,y)=(0,0)$ coordinates, and then subjected to another round of filtering by temporal length, by the ratio of position samples in branches (further eliminating bias from overfitting one of the branches), and another stage of multiple turning point instance elimination. Finally, the misalignment of the orientation of the U-turn with the alternating MF axis is checked -- the U-turns and the MF axis are rotated to align the MF along the $x$ axis, and then, for all the U-turns, the ratio of the coordinate sums of $x$ and $y$ over both branches of the U-turn is computed. If the ratio is too low, the U-turn does not correspond to a motion compliant with the MF. The remaining U-turns are eligible for fitting to \eqref{eq:uturn-shape-equation-final}.

To fit U-turn coordinate sets to \eqref{eq:uturn-shape-equation-final}, U-turn turning points, previously translated to initial coordinates $(0,0)$, are used as anchors for shape fitting. The U-turn is fitted to the offset version of the theoretical shape function \eqref{eq:uturn-shape-equation-final} $y_\text{offset}(x,L,x_0,y_0) = y_0 + y(x-x_0,L)$ given by the shape function translation $(x_0,y_0)$ and the parameter value $L$ by finding the optimal U-turn coordinate set $\vec{r}_k (\varphi) = (x_k,y_k),~k \in \mathbb{N}$ generated by rotating the initial U-turn by the angle $\varphi$, such that the following optimization problem is solved:

\begin{equation}
   L: \arg \min_{(x_0,y_0,\varphi,L)} f(x_0,y_0,\varphi,L) + \beta \cdot g (\varphi)
   \label{eq:optimization-fitness-total}
\end{equation}
where $\beta \geq 0,~ L>0,~ \varphi \in [0;2 \pi),~ x_0 \in I_1 (\vec{r}_k),~ y_0 \in I_2 (\vec{r}_k)$. The functions $f$ and $g$ are defined as follows:

\begin{equation}
    f(x_0,y_0,\varphi,L) = \frac{1}{\sum_k \norm{\vec{r}_k (\varphi)}} \sum_k \left(y_\text{offset}(x_k(\varphi),L,x_0,y_0) - y_k(\varphi) \right)^2
    \label{eq:optimization-fitness-main}
\end{equation}

\begin{equation}
    g(\varphi) = \frac{1}{\sum_k y_k^2(\varphi)} \sum_{y_k(\varphi)>0} y_k^2(\varphi)
    \label{eq:optimization-penalty-function}
\end{equation}

The solution minimizes the discrepancy between the U-turn data points and the theoretical shape function \eqref{eq:optimization-fitness-main} while penalizing U-turn rotations resulting in U-turn points with $y_k>0$ \eqref{eq:optimization-penalty-function}. The latter is implemented because \eqref{eq:uturn-shape-equation-final} has the property $y(x,L) < 0~ \forall x \neq 0$. Furthermore, to enforce $f+\beta g: \mathbb{R} \rightarrow \mathbb{R}~ \forall \{ x_k(\varphi), L \}$, the $y_\text{offset}(x_k(\varphi),L,x_0,y_0)$ values are sampled for every $L$ instance $N_{\varepsilon}$ times uniformly in the $x_k \in [ -L + \varepsilon; L - \varepsilon]$ interval. This is due to $y(x,L) \in \mathbb{C}~ \forall \abs{x}>L$. As a result, $\left(y_\text{offset}(x_k(\varphi),L,x_0,y_0) - y_k(\varphi) \right)^2$ in \eqref{eq:optimization-fitness-main} are instead computed as

\begin{equation}
   \min_{s \in \mathbb{N}} \norm{\vec{r}_s - \vec{r}_k (\varphi)}^2;~ \vec{r}_s = (y_s,x_s): y_s = y_\text{offset}(x_s,L,x_0,y_0),~ x_s \in [ -L + \varepsilon; L - \varepsilon]
\end{equation}
where $\vec{r}_s$ are the sampling points for $y_\text{offset}$. Note that \eqref{eq:optimization-fitness-total} penalizes temporally and spatially longer tracks slightly less, which is intentional -- this way, better reconstructed U-turns are prioritized.

After all U-turn fits are complete and $L$ determined, the $L$ values and the U-turn data are filtered by their $f + \beta g$ and $g$ values. Then $\alpha$ is computed from MTB and flow medium properties assuming spherically shaped MTB:

\begin{equation}
    \alpha = 8 \pi \eta R^3
    \label{eq:mtb-drag-shpherical}
\end{equation}
Note, however, that any other rotational drag model can be used based on the morphology of cells -- for example, cylindrical \cite{popp_polarity_2014}, ellipsoidal \cite{leao_eliptic_drag_c}, the sphere chain model presented in \cite{mtb-rotational-drag-sphere-chain-model}, or an experimental spirillum drag model as in \cite{Pichel_2018}.

We then estimate the characteristic U-turn $v$ to compute $m$ from \eqref{eq:moment-from-uturn-params}. This is done by computing the U-turn curvature over the arc length and computing the mean velocity over the segment with curvature exceeding a relative threshold, that is, velocity is not sampled from the straighter segments of the track.

All of the above procedures and the parameters are summarized in Algorithm \ref{alg:mtb-moment-extraction} in Appendix \ref{appendix:moment-calculation-algorithm-parameters}, and are also provided with the code, which is linked in the Code Availability section.

\section{Results}
\label{sec:results}

\begin{figure}[h!]
\begin{center}
\includegraphics[width=1.0\textwidth]{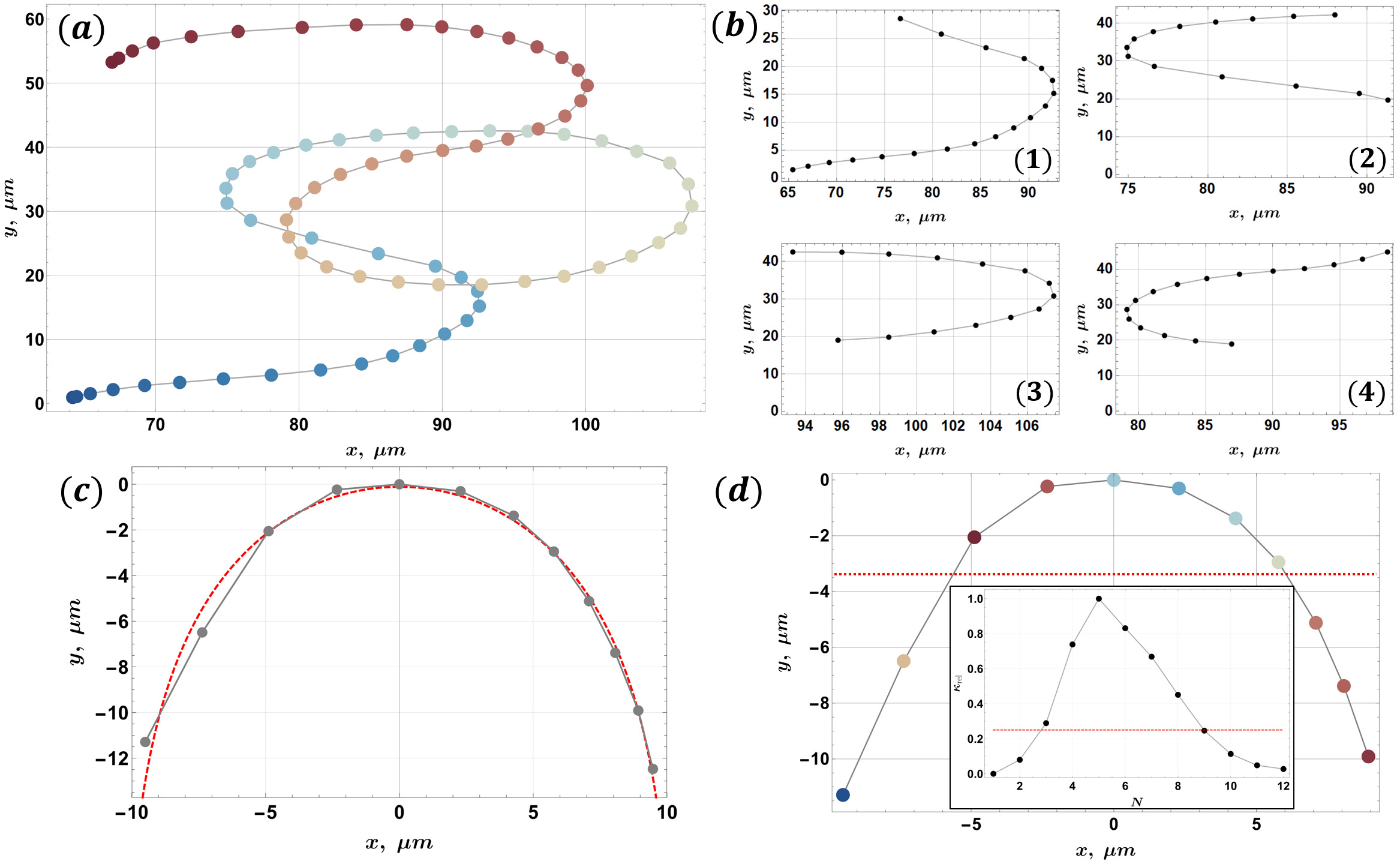} 
\caption{An example of U-turn decomposition and MTB moment calculation: (a) a MSR-1 track in an alternating MF, where points are color coded chronologically from blue to red; (b) track (a) decomposed into individual U-turns (1-4), before symmetrization/translation (4 out of 5 detected U-turns shown); (c) U-turn (4) filtered and fitted to the \eqref{eq:uturn-shape-equation-final}; (d) evaluating U-turn velocity (color-coded by magnitude from blue to red) $v$ for \eqref{eq:moment-from-uturn-params}: U-turn $v$ is computed as the mean for the U-turn interval where the relative curvature (inset) is above a threshold. Please see Algorithm \ref{alg:mtb-moment-extraction} for details and parameters.}
\label{fig:u-turn-decomposition}
\end{center}
\end{figure}

\begin{figure}[h!]
\begin{center}
\includegraphics[width=0.6\textwidth]{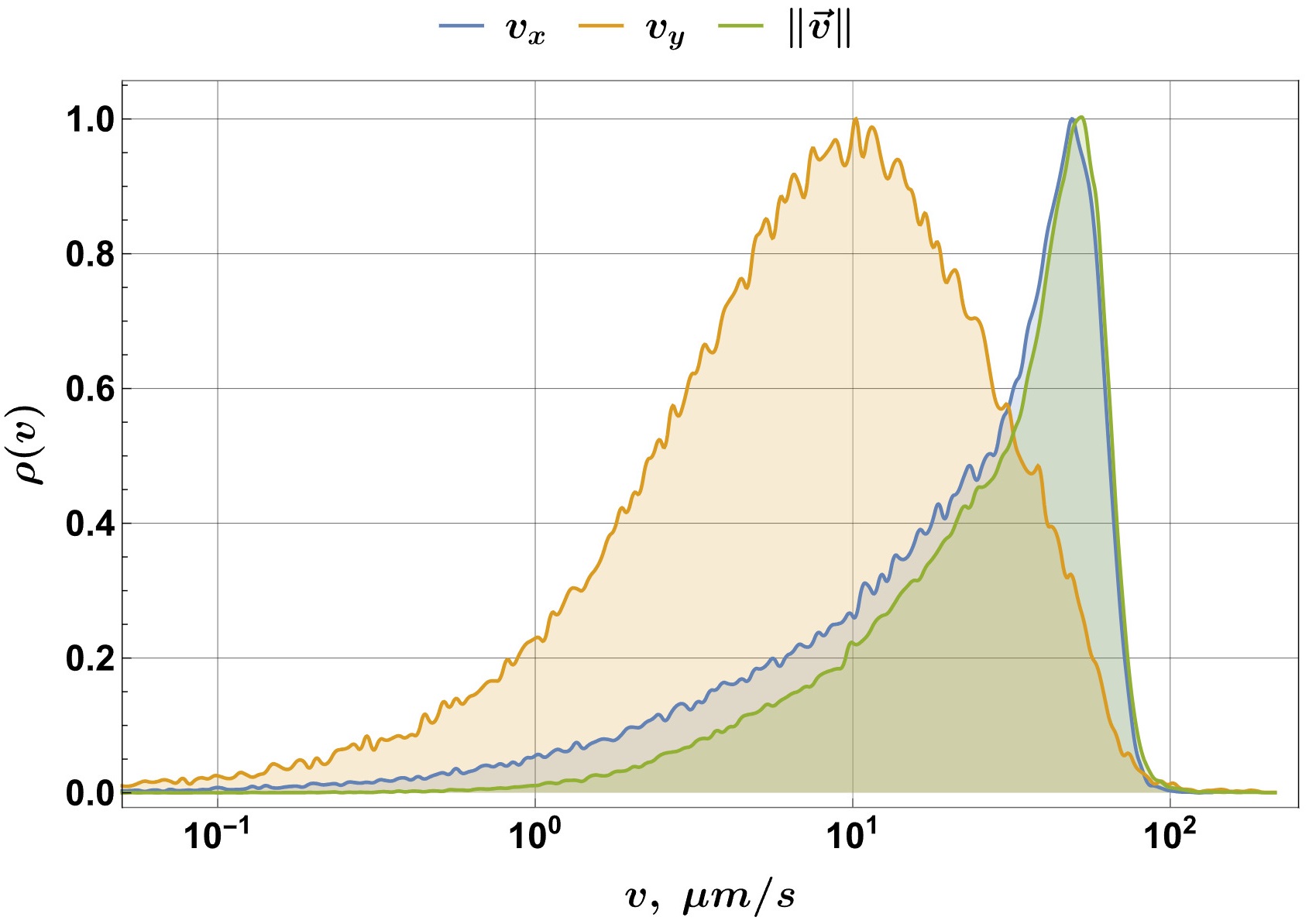} 
\caption{Normalized probability $\rho$ for the MSR-1 MTB velocity with MF applied in the $x$ direction (horizontal image axis). The most probable MTB velocity magnitude is $\norm{\vec{v}} \sim 51.3~ \mu m/s$.}
\label{fig:velocity_histogram}
\end{center}
\end{figure}

\begin{figure}[h!]
\begin{center}
\includegraphics[width=0.8\textwidth]{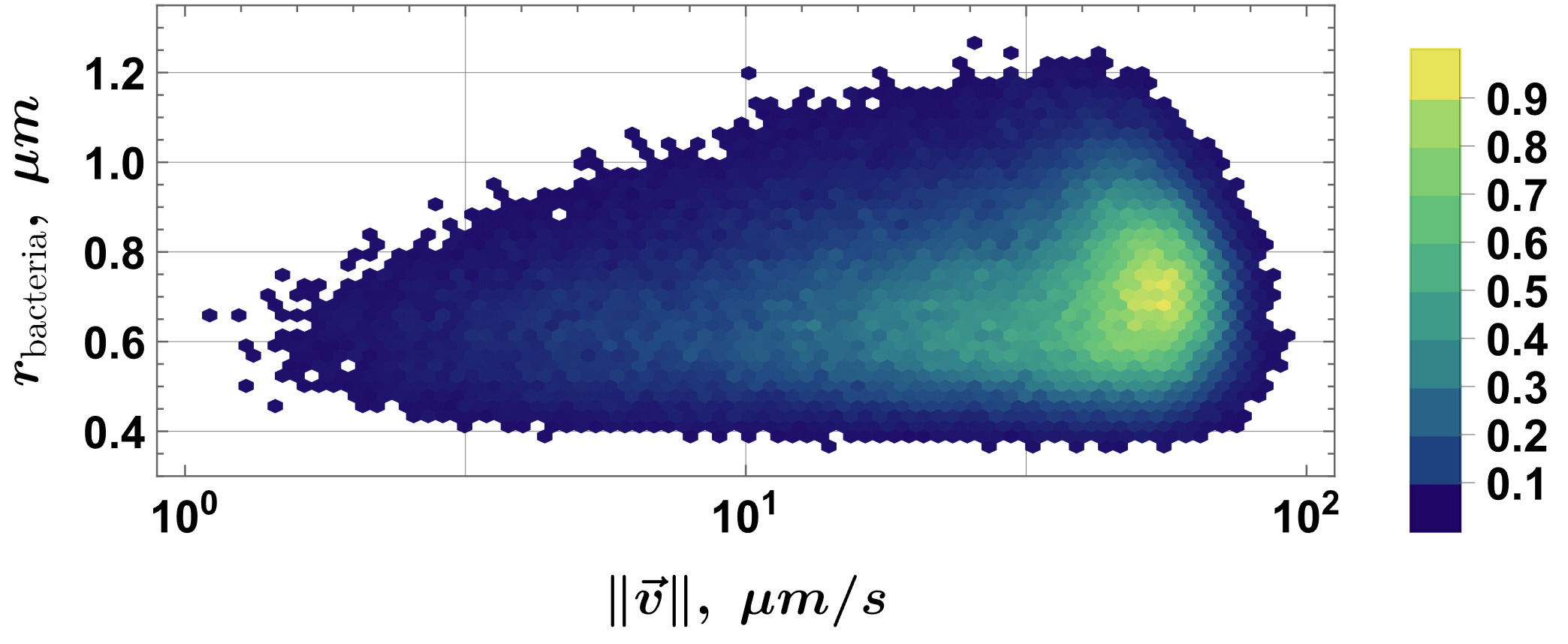} 
\caption{A histogram relating the effective radii $r$ measured for the cell population and cell velocity magnitude $\norm{\vec{v}}$. The relative bin size is $0.015$ of the data range, and value display relative  threshold $0.02$.}
\label{fig:velocity_radius_histogram_hexplot}
\end{center}
\end{figure}

\begin{figure}[h!]
\begin{center}
\includegraphics[width=0.70\textwidth]{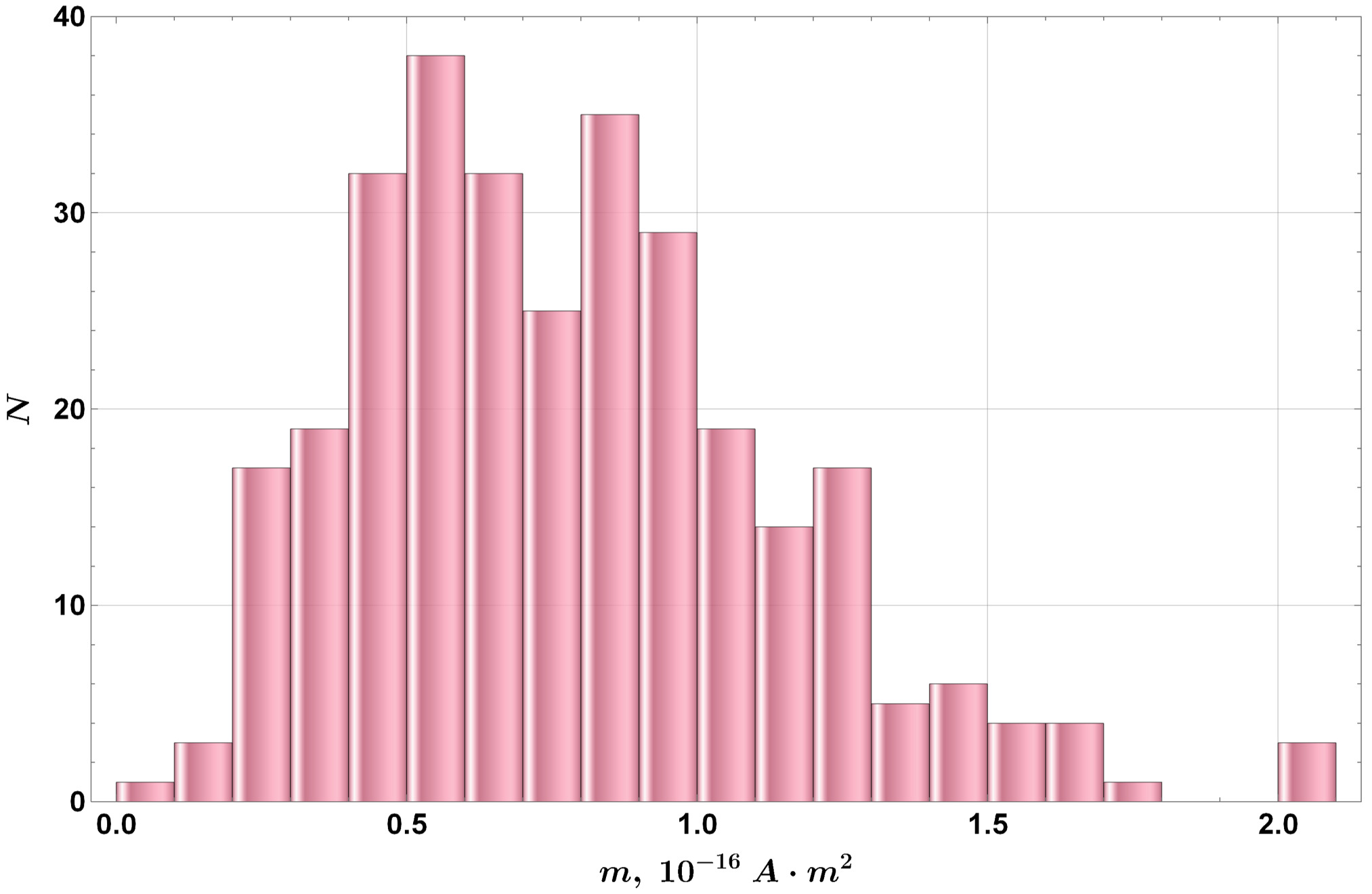} 
\caption{MSR-1 magnetic moment histogram (count $N$ versus moment $m$, Freedman-Diaconis binning), calculated using the proposed U-turn shape function \eqref{eq:uturn-shape-equation-final} and U-turn analysis algorithm (Algorithm \ref{alg:mtb-moment-extraction}), $N_\text{total}=304$.}
\label{fig:m_moment_histogram}
\end{center}
\end{figure}

\begin{figure}[h!]
\begin{center}
\includegraphics[width=0.85\textwidth]{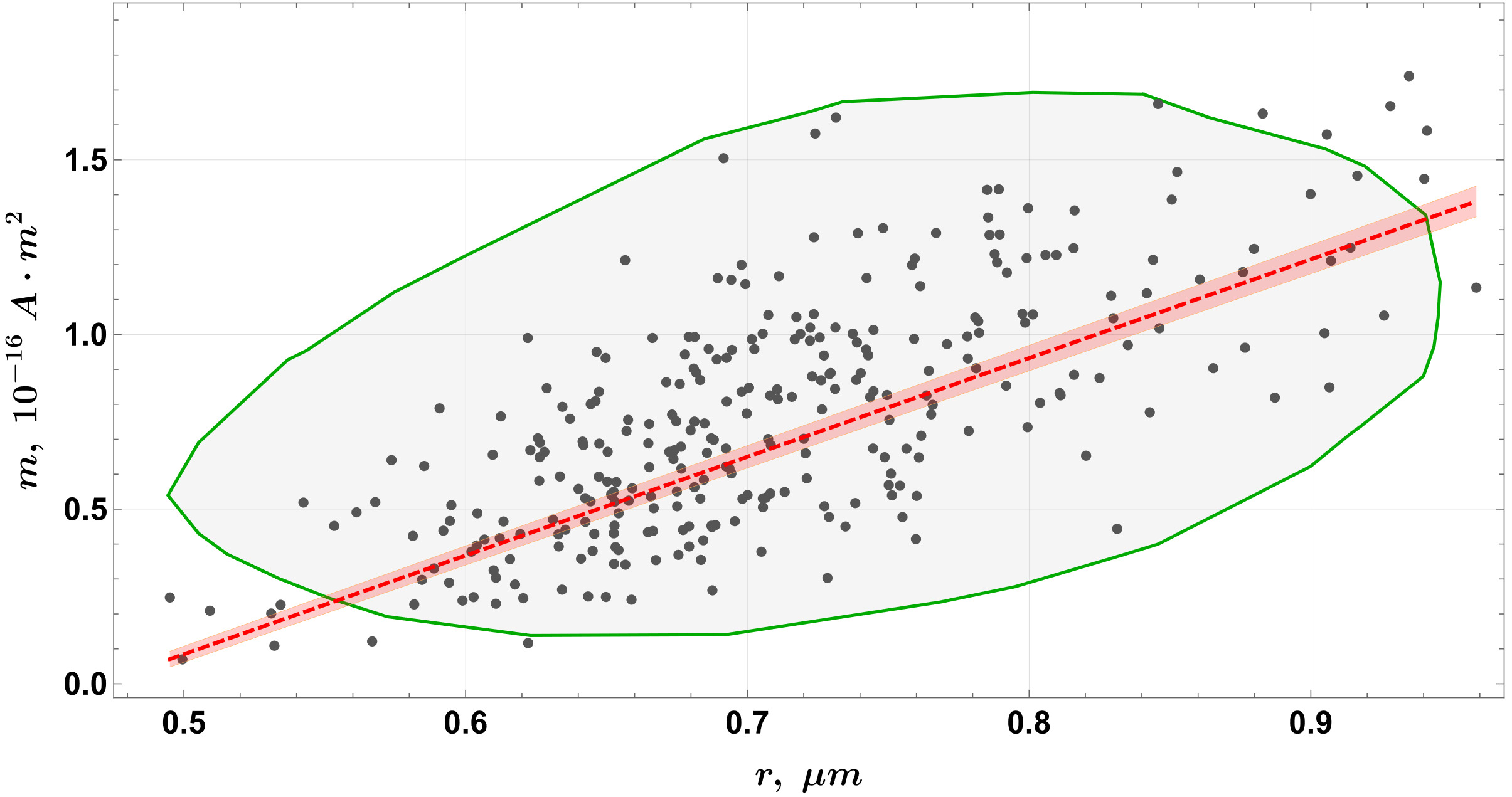}\caption{Magnetic moment $m$ versus MTB effective radius $r$: 304 values determined using the proposed U-turn shape function \eqref{eq:uturn-shape-equation-final} and Algorithm \ref{alg:mtb-moment-extraction} are represented by gray dots, with the $q=0.95$ quantile uncertainty region indicated as the light gray area with a green boundary, and the error-weighed $m(r) = C_1 + C_2 \cdot r$ linear fit given by the dashed red line with light red slope error margins ($R^2 = 0.611$). Here $C_1 = (2.83 \pm 0.13) \cdot 10^{-16}~ A\cdot m^2$ and $C_2 = - (1.33 \pm 0.09) \cdot 10^{-10}~ A\cdot m^3$, with $m(r_c)=0$ for $r_c = 0.47 ~\mu m$. Fit weights are inverse squares of the total uncertainty due to $m$ and $r$ uncertainties (error propagation). The uncertainty region is derived by computing 250 directional quantiles \cite{quantile-tomography, antonov-directional-quantile-envelopes}, connecting the quantiles with a smooth curve, and applying curve evolution polyline simplification \cite{polyline-simplification-curve-evolution} to the resulting quantile envelope.}
\label{fig:moment_vs_radius}
\end{center}
\end{figure}

We applied our new method to calculate $m$ via the U-turn method for a sample population of MTB. We imaged the motion of MSR-1 MTB in alternating MF using an optical microscope (Section \ref{sec:microscopy-image-acquisition}) and then recovered the cell trajectories from the acquired images (Section \ref{sec:detection-tracking}). We then analyzed $\sim 18~\text{K}$ trajectories retrieved from $1~\text{K}$ images that yielded 304 eligible U turns (i.e., from trajectories and subsequently U-turns that survived the filtering stages in Algorithm \ref{alg:mtb-moment-extraction}) for which $m$ was calculated (Sections \ref{sec:theory-mtb-motion-model} and \ref{sec:determining-L}). An example of a U-turn segmentation and processing from a trajectory is shown in Figure \ref{fig:u-turn-decomposition}), and the population cell radii and velocity can be seen in Figures \ref{fig:velocity_histogram} and \ref{fig:velocity_radius_histogram_hexplot}.
The resulting overall $m$ statistics are shown in Figure \ref{fig:m_moment_histogram},
and the relationship between $m$ and the effective cell radius $r$ is presented in Figure \ref{fig:moment_vs_radius}. $m$ data obtained using the proposed theoretical U-turn shape function \eqref{eq:uturn-shape-equation-final} were then compared with the results obtained using the U-turn time $\tau$-based calculations via \eqref{eq:magnetic-moment-tau} (Figures \ref{fig:tau_moment_histogram} and \ref{fig:moment_vs_radius_tau}). Figures \ref{fig:moment_histogram_comparison_L_vs_tau} and \ref{fig:moment_vs_radius_comparison_L_vs_tau} provide a direct visual comparison.

Furthermore, we tested the sensitivity of our proposed method with respect to different estimates of the U-turn velocity $v$ for the $m$ equation \eqref{eq:moment-from-uturn-params}. $m$ statistics for three $v$ estimates are compared in Figure \ref{fig:moment_diff_velocity_estimates}. The simplest assumes that the U-turn $v$ is the most probable population velocity, determined from the velocity statistics derived from the trajectory data, as seen in Figures \ref{fig:velocity_histogram} and \ref{fig:velocity_radius_histogram_hexplot}. The maximum probability population velocity is $51.3 ~\mu m/s$ with a standard deviation $\sim 20.3 ~\mu m/s$, which is consistent with the results reported in \cite{Pichel_2018}, but differs from the $30 ~\mu m/s$ reported in \cite{swimming-organism-data-bank}, although it is arguably within the error margin. The other estimates are obtained via our proposed method (Algorithm \ref{alg:mtb-moment-extraction}), which estimates the value of $v$ for each U-turn as seen in Figure \ref{fig:u-turn-decomposition} (and Section \ref{sec:determining-L}, which also illustrates the U-turn decomposition process. We use both the individual $v$ and the mean U-turn $v$ (not to be confused with the entire population mean).

\begin{figure}[h!]
\begin{center}
\includegraphics[width=0.85\textwidth]{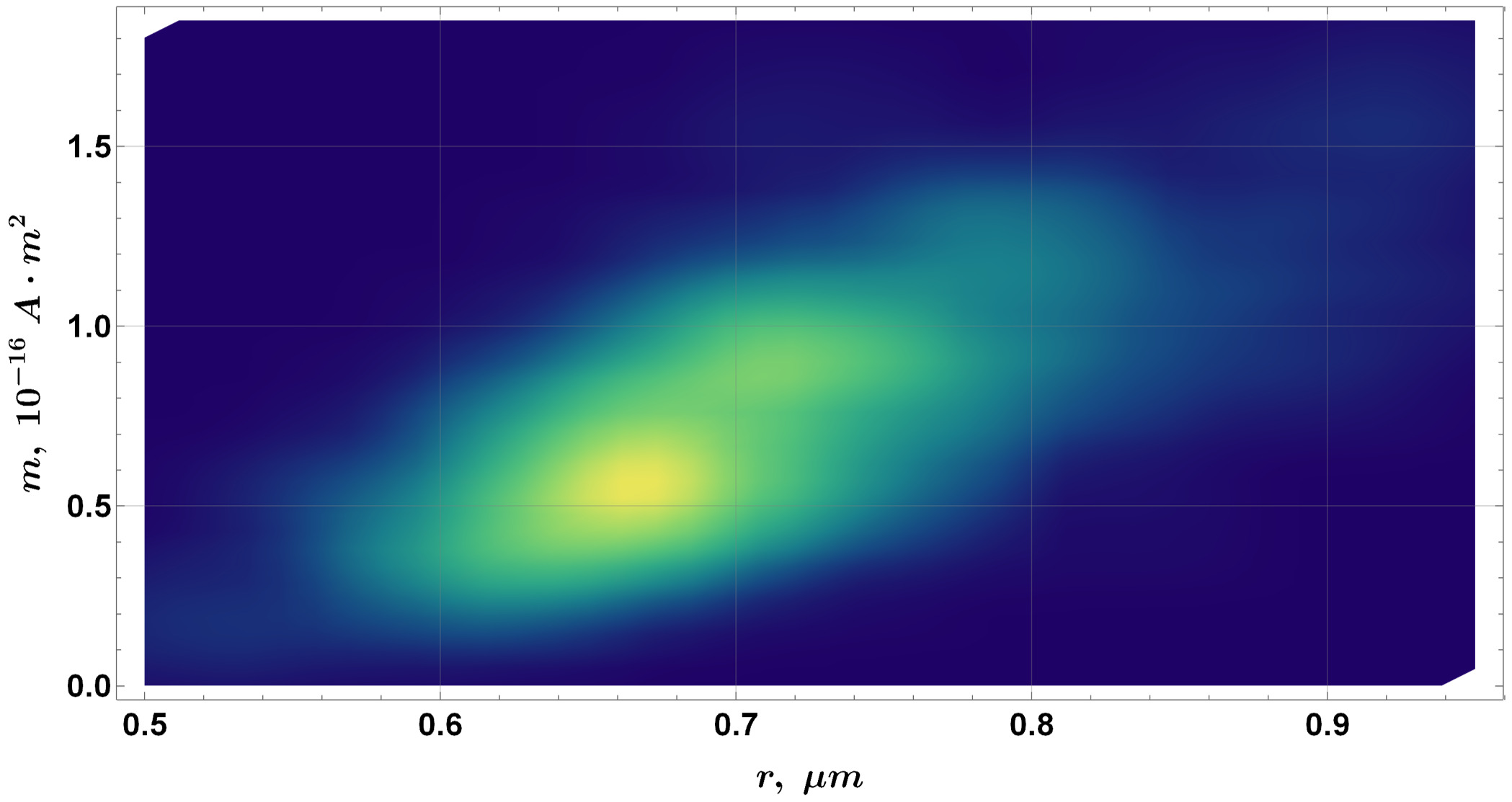} 
\caption{Magnetic moment $m$ versus MTB effective radius $r$: a smooth density histogram (Sheather-Jones bandwidth estimator, Epanechnikov kernel) computed for $m$ determined via proposed U-turn shape function \eqref{eq:uturn-shape-equation-final} and Algorithm \ref{alg:mtb-moment-extraction}.}
\label{fig:moment_vs_radius_smooth_histogram}
\end{center}
\end{figure}

\begin{figure}[h!]
\begin{center}
\includegraphics[width=0.70\textwidth]{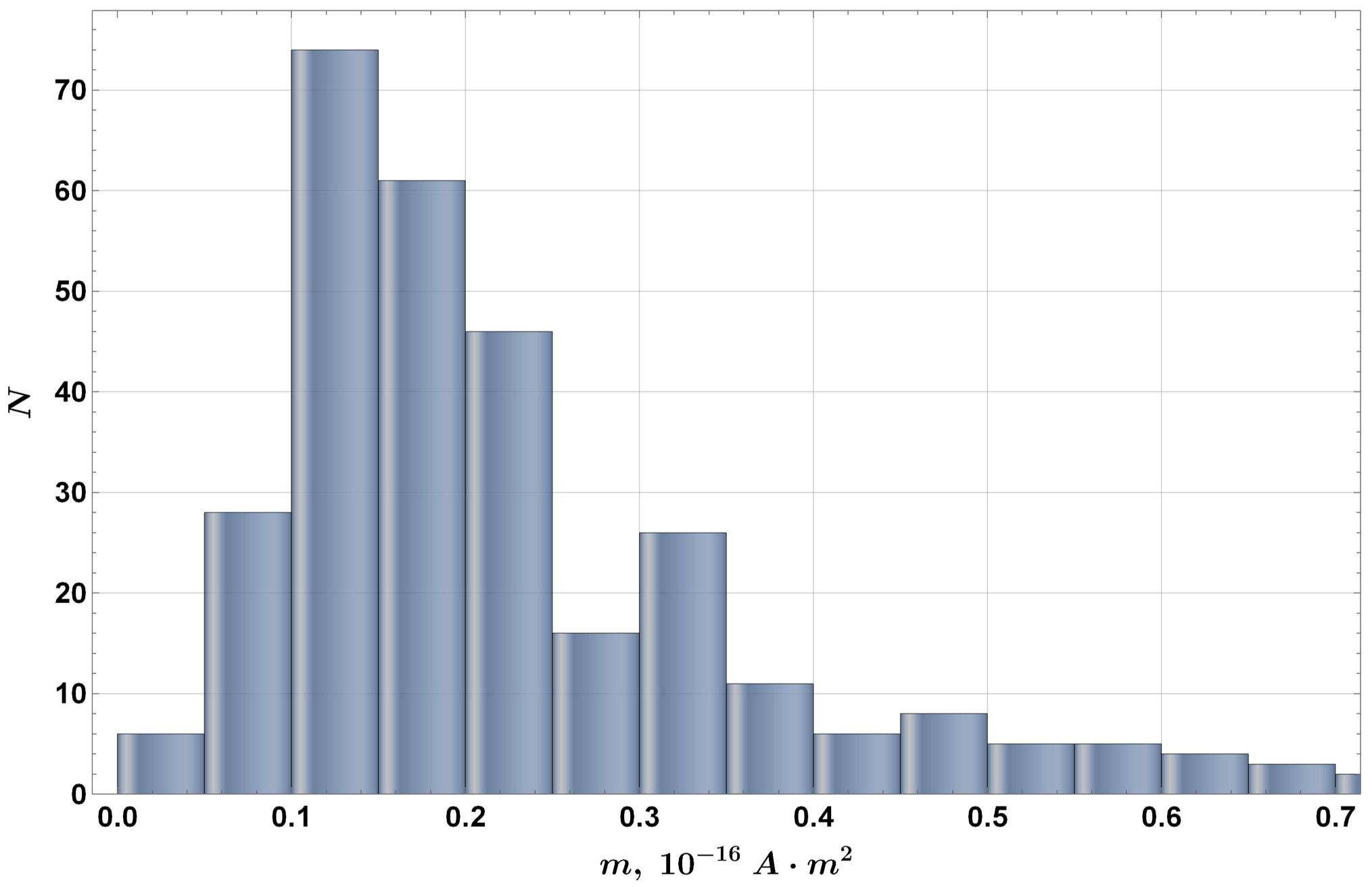} 
\caption{MSR-1 magnetic moment histogram (count $N$ versus moment $m$, Freedman-Diaconis binning), calculated using the existing U-turn time $\tau$-based formula \eqref{eq:magnetic-moment-tau} and the proposed U-turn decomposition algorithm (Algorithm \ref{alg:mtb-moment-extraction}), $N_\text{total}=264$.}
\label{fig:tau_moment_histogram}
\end{center}
\end{figure}

\begin{figure}[h!]
\begin{center}
\includegraphics[width=0.85\textwidth]{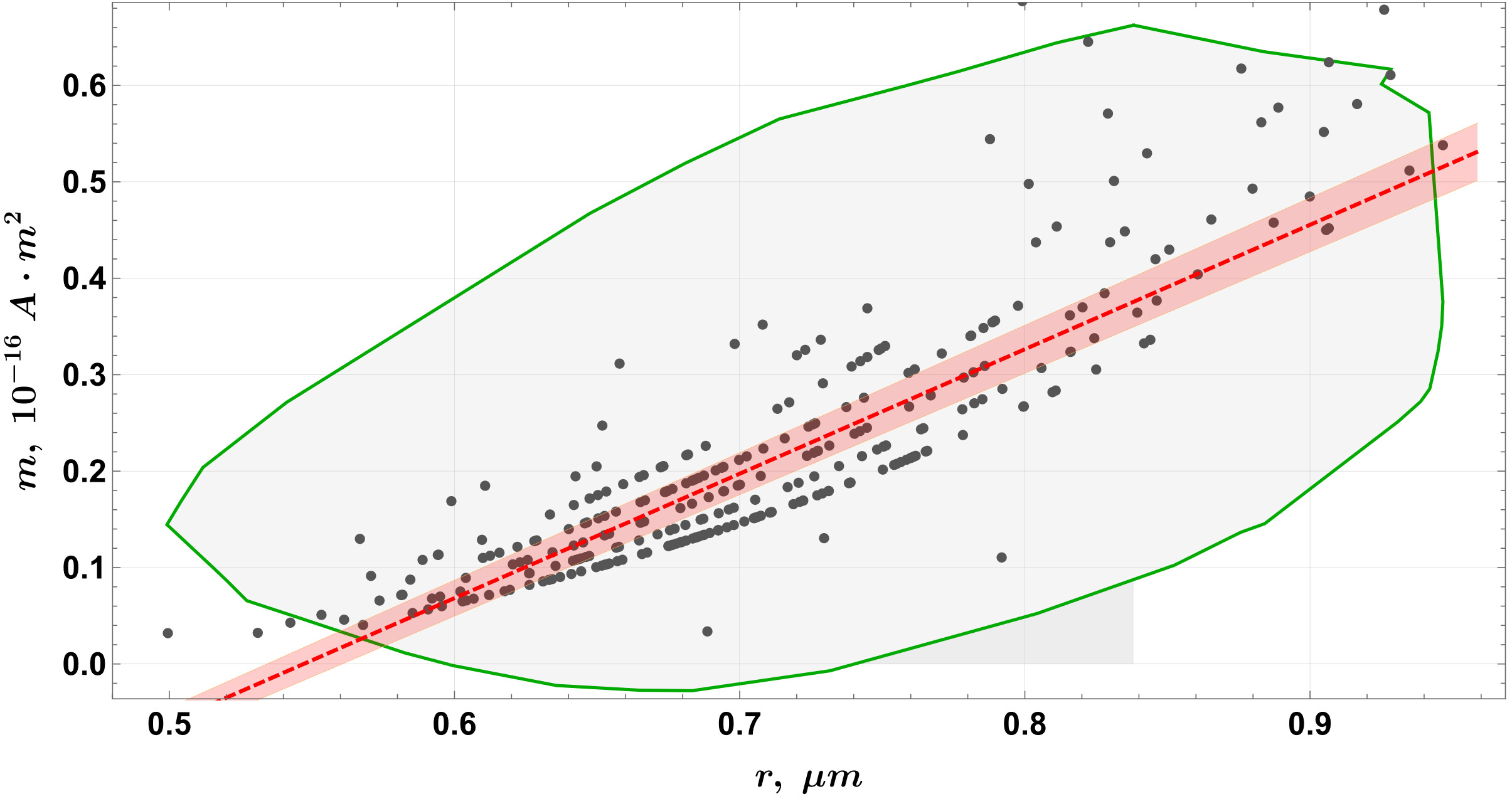}\caption{Magnetic moment $m$ versus MTB effective radius $r$: 264 values determined using the U-turn time $\tau$-based formula \eqref{eq:magnetic-moment-tau} and the proposed U-turn decomposition (Algorithm \ref{alg:mtb-moment-extraction}), with a $m(r) = C_1 + C_2 \cdot r$ linear fit ($R^2 = 0.773$). Here $C_1 = (1.290 \pm 0.129) \cdot 10^{-16}~ A\cdot m^2$ and $C_2 = - (0.706 \pm 0.086) \cdot 10^{-10}~ A\cdot m^3$, with $m(r_c)=0$ for $r_c = 0.55 ~\mu m$.}
\label{fig:moment_vs_radius_tau}
\end{center}
\end{figure}

Using our proposed $m$ calculation algorithm and the individual estimate of the U turn $v$ (Figure \ref{fig:u-turn-decomposition}), we obtained the $m$ histogram as seen in Figure \ref{fig:m_moment_histogram} where most cells have $m \in (0.4; 1.0)  \cdot 10^{-16}~ A \cdot m^2$. However, viewing $m$ statistics this way only does not provide the full picture -- therefore, $m$ scaling with $r$ is presented in Figure \ref{fig:moment_vs_radius}. Note the positive correlation between the cell $m$ and $r$ values. To assess how $m$ scales with $r$, the data is assigned an error-weighed linear fit $m(r) = C_1 + C_2 \cdot r$, where $C_1 = dm/dr = (2.83 \pm 0.13) \cdot 10^{-16}~ A\cdot m^2$ and $C_2 = - (1.33 \pm 0.09) \cdot 10^{-10}~ A\cdot m^3$. This implies a critical $r=r_c$ for which cell size is insufficient to host magnetosomes that generate a significant $m$; here $m(r_c)=0$ for $r_c \sim 0.47 ~\mu m$. The linear fit slope in Figure \ref{fig:moment_vs_radius} is also quite well aligned with the inclination of the quantile uncertainty region $q=0.95$. In Figure \ref{fig:moment_vs_radius_smooth_histogram} it can be observed that the most probable effective radii among the MTB corresponding to eligible U-turns are $r \sim 0.66~ \mu m$ with $m \sim 0.6~ \cdot 10^{-16}~ A \cdot m^2$, and $r \sim 0.72~ \mu m$ with $m \sim 0.9~ \cdot 10^{-16}~ A \cdot m^2$, respectively.

To determine the U-turn $\tau$ values (the characteristic U-turn completion time) and compare the values of $m$ between the standard equation \eqref{eq:magnetic-moment-tau} and the proposed ones, \eqref{eq:uturn-shape-equation-final} and \eqref{eq:moment-from-uturn-params}, we compute $\tau$ and $m(\tau)$ for the same U-turns we used to estimate $m$ based on the proposed approach. Using the MF switching data, we detect the MF switch timings from the coil voltage time series by thresholding the time series with the $0.95$ relative voltage values, and finding the nonzero differences in the resulting signal. We then match the MF switch timings to the U-turn timestamps and find the intersection between the two MF switch timestamps nearest to the U-turn temporal endpoints, and the U-turn time interval. The $\tau$ values are then the temporal lengths of the U-turns truncated by the MF switch timestamps multiplied by the image acquisition frame duration. The resulting statistics based on $\tau$ based $m$ are shown in Figure \ref{fig:tau_moment_histogram}, where most cells have $m \in (0.05;0.25) \cdot 10^{-16}~ A \cdot m^2$. The $m$ scaling with $r$ is shown in Figure \ref{fig:moment_vs_radius_tau}, where the experimental limitation clearly appears when using $\tau$ as a parameter that characterizes the U-turn motion of the cell -- its precision is strongly limited by the frame rate used for image acquisition. As a result, Figure \ref{fig:moment_vs_radius_tau} shows banding artifacts in the $m(r)$ dataset computed from $\tau$. This is because the $\tau$ values are multiples of the frame duration, and therefore the differences in $m(r)$ are in the form of clearly visible steps and distinct curves. As frame rate tends to infinity, these artifacts should disappear, but for lower frame rates the effect is very significant. While $L$ is also limited by the frame rate, $L \in \mathbb{R}$, which means that the continuum of $(m,r)$ values is filled much more uniformly by sampling the MTB population $m$ values from $L$ values for U-turns. The calculation of $m$ through $L$ is more limited by image resolution, which in this case is $1920 \times 1200~ {px}^2$ ($\sim 28~ px$ per MTB length), which is usually less of an issue than a lower frame rate.

Importantly, Figure \ref{fig:moment_vs_radius_tau} is not the first instance in which the $m(r)$ data point banding can be observed -- $m(r)$ correlation was also considered in \cite{alvaros_u_turn_cocci}, although for a wild sample of magnetic cocci, not MSR-1. Instead of $r$, the $m$ versus the MTB volume $V$ correlation was produced (\cite{alvaros_u_turn_cocci}[Figure 10]), which exhibited banding within the plot region with a higher data point number density. We therefore argue that our observations are not a coincidence. Additionally, the frame rate in \cite{alvaros_u_turn_cocci} was $82$ FPS (versus our $20$ FPS) while the most probable MTB population velocity was estimated at $\sim 80 \mu m/s$ (\cite{alvaros_u_turn_cocci}[Figure 6] versus our $\sim 51.3~ \mu m/s$), meaning that the corresponding temporal sampling should be better than in our case by a factor of $\sim 2.6$, but the artifacts in the $m(V)$ correlation are still present.

\begin{figure}[h!]
\begin{center}
\includegraphics[width=0.85\textwidth]{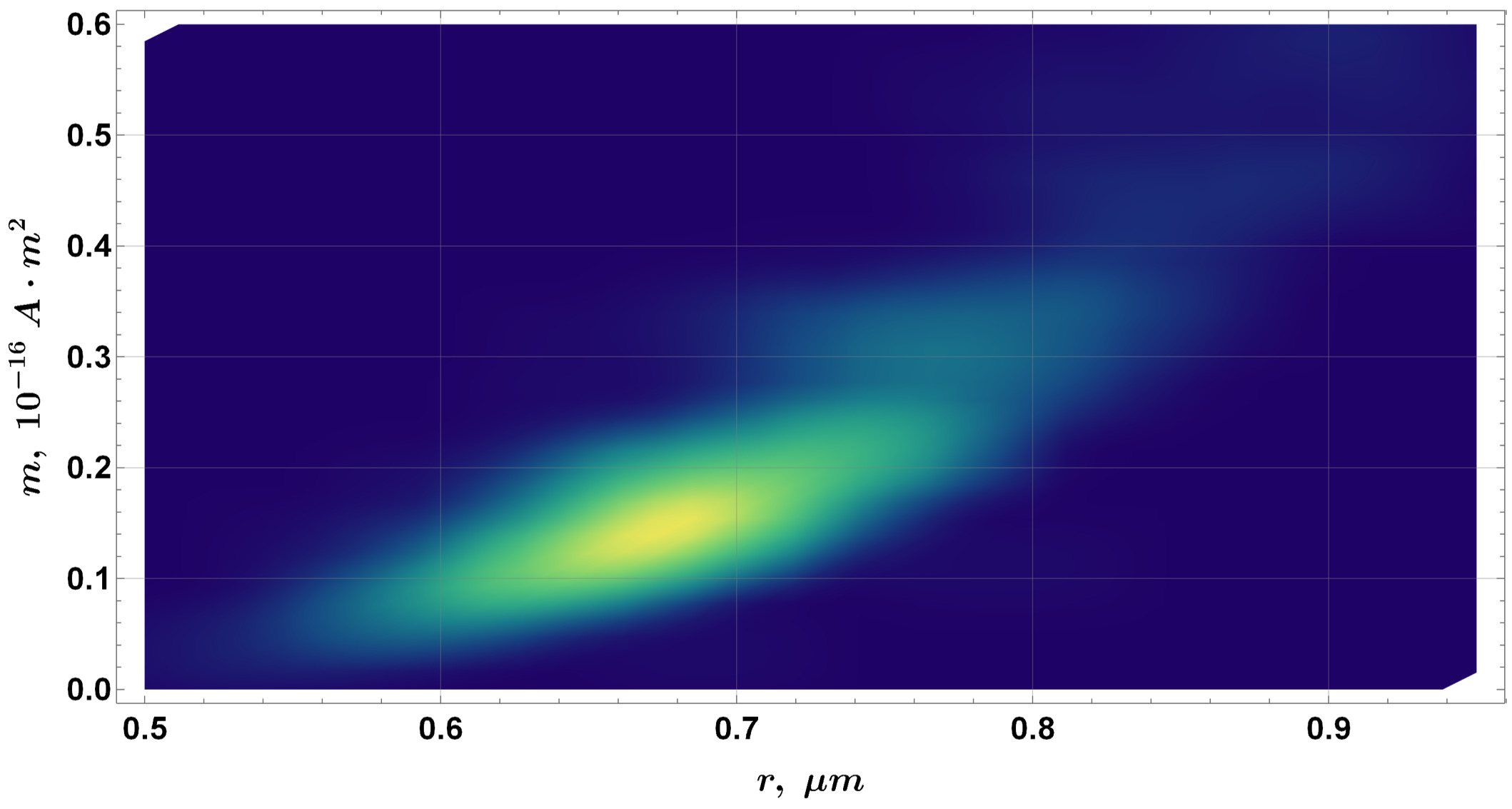} 
\caption{Magnetic moment $m$ versus MTB effective radius $r$: a smooth density histogram (Sheather-Jones bandwidth estimator, Epanechnikov kernel) computed for $m$ determined via the U-turn time $\tau$-based formula \eqref{eq:uturn-shape-equation-final} and U-turn decomposition proposed in Algorithm \ref{alg:mtb-moment-extraction}.}
\label{fig:moment_vs_radius_smooth_histogram_tau}
\end{center}
\end{figure}

\begin{figure}[h!]
\begin{center}
\includegraphics[width=0.85\textwidth]{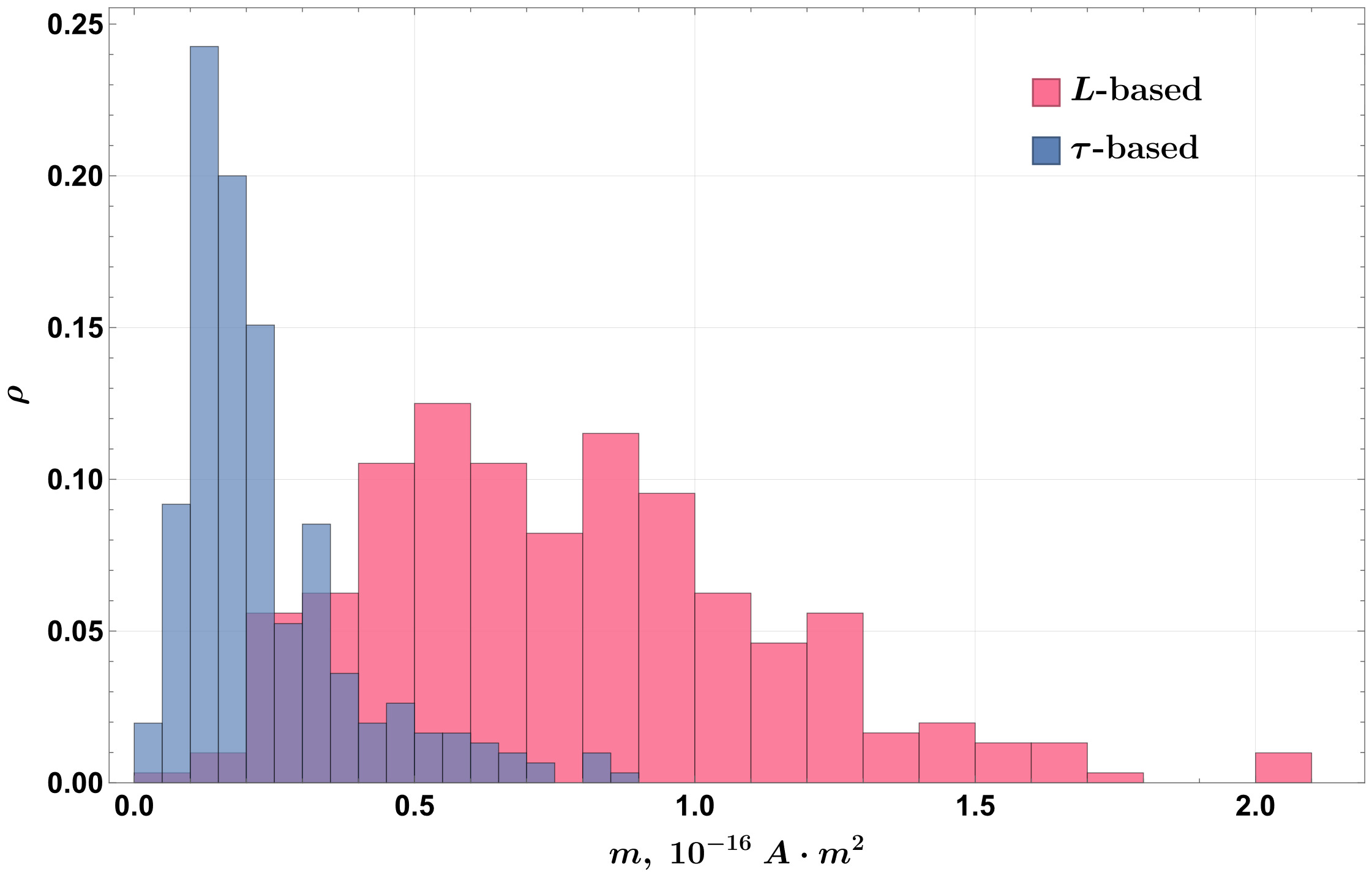} 
\caption{MSR-1 magnetic moment $m$ probability $\rho$ histograms (Freedman-Diaconis binning) for $m$ calculated using the proposed U-turn shape function \eqref{eq:uturn-shape-equation-final} and U-turn analysis algorithm (Algorithm \ref{alg:mtb-moment-extraction}), colored blue, and using the existing U-turn time $\tau$-based formula \eqref{eq:magnetic-moment-tau} and the proposed U-turn decomposition algorithm (Algorithm \ref{alg:mtb-moment-extraction}), colored light-red.}
\label{fig:moment_histogram_comparison_L_vs_tau}
\end{center}
\end{figure}

\begin{figure}[h!]
\begin{center}
\includegraphics[width=0.85\textwidth]{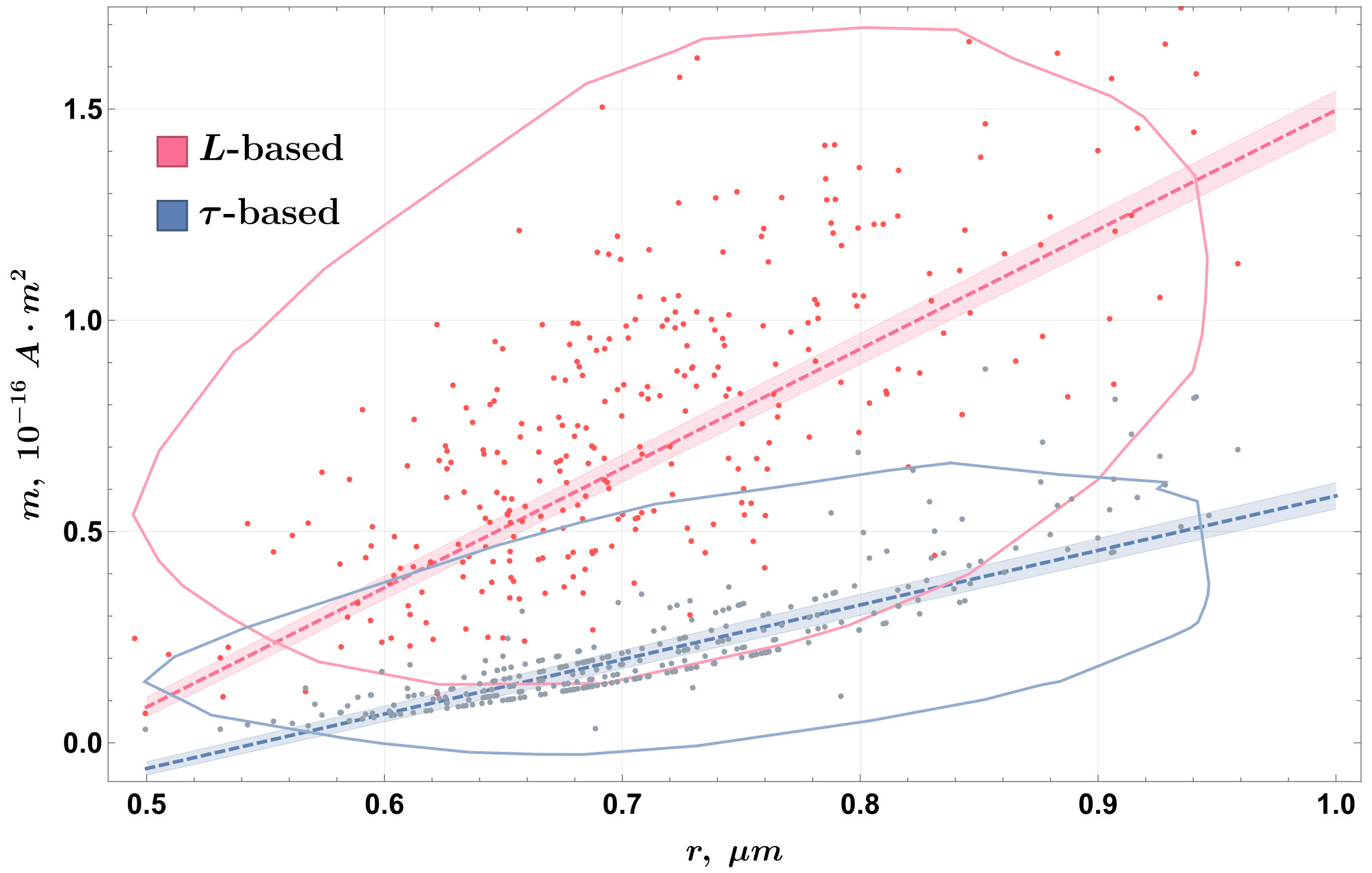}\caption{Magnetic moment $m$ versus MTB effective radius $r$: linear fits and $q=0.95$ uncertainty regions \eqref{eq:uturn-shape-equation-final} for $m$ calculated using the proposed U-turn shape function \eqref{eq:uturn-shape-equation-final} and U-turn analysis algorithm (Algorithm \ref{alg:mtb-moment-extraction}), colored blue, and using the existing U-turn time $\tau$-based formula \eqref{eq:magnetic-moment-tau} and the proposed U-turn decomposition algorithm (Algorithm \ref{alg:mtb-moment-extraction}), colored light-red.}
\label{fig:moment_vs_radius_comparison_L_vs_tau}
\end{center}
\end{figure}

\begin{figure}[h!]
\begin{center}
\includegraphics[width=0.85\textwidth]{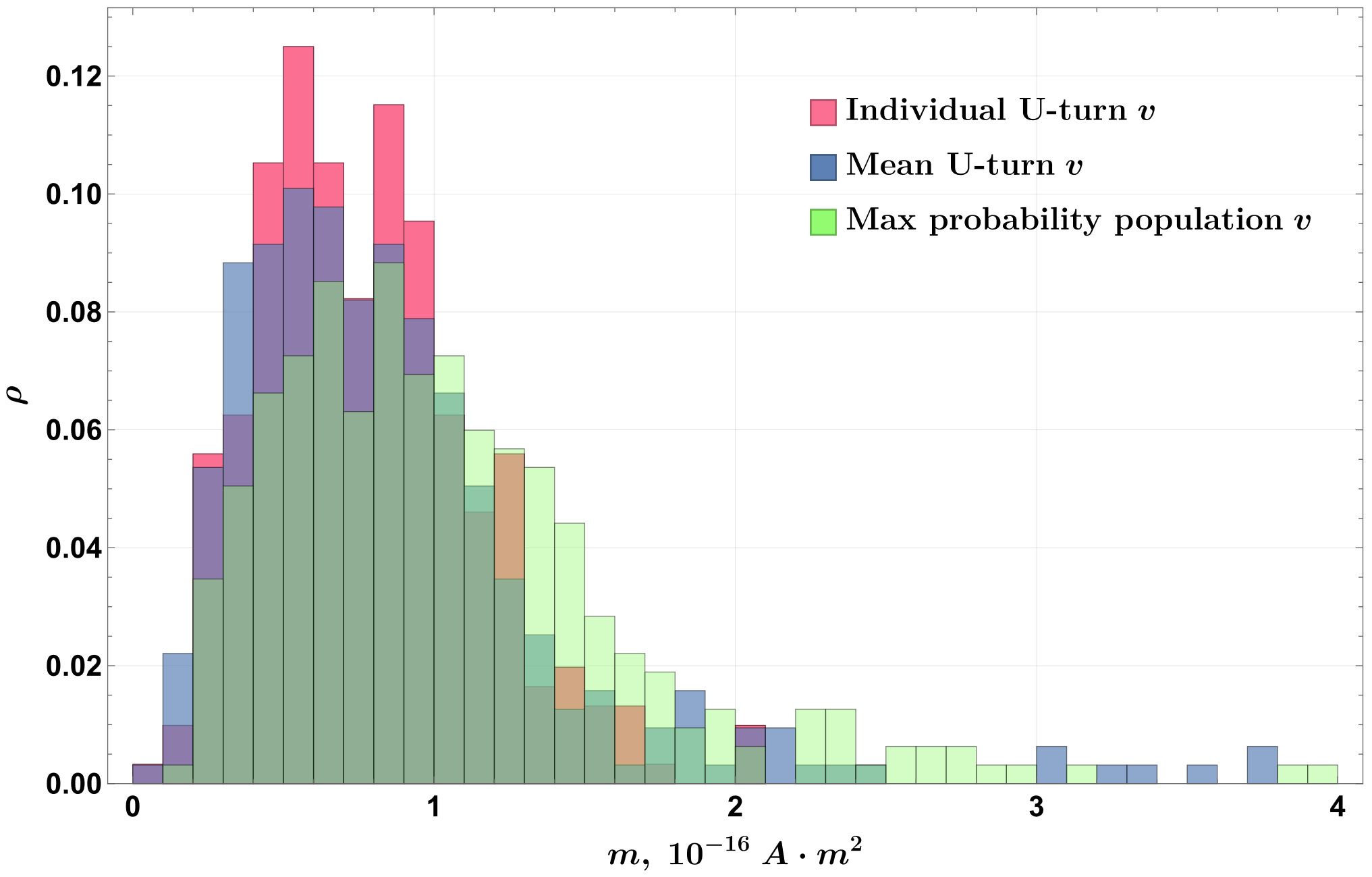}\caption{MSR-1 magnetic moment $m$ probability histograms (Freedman-Diaconis binning) for $m$ calculated using the proposed U-turn shape function \eqref{eq:uturn-shape-equation-final} and U-turn analysis algorithm (Algorithm \ref{alg:mtb-moment-extraction}). Characteristic velocity used in \eqref{eq:uturn-shape-equation-final}: individual U-turn values (light-red), mean value for all U-turns (blue) and the most probable velocity magnitude for the observed MTB population (light-green; Figure \ref{fig:velocity_histogram}).}
\label{fig:moment_diff_velocity_estimates}
\end{center}
\end{figure}

Similarly to the $m(L)$ case, a trend in $m(\tau)$ can be observed where $m$ increases with $r$, with the most probable MTB $r$ among the considered U-turns $r \sim 0.68~ \mu m$ with $m \sim 0.15~ \cdot 10^{-16}~ A \cdot m^2$, as seen in Figure \ref{fig:moment_vs_radius_smooth_histogram_tau}. Although the $m(r)$ data for $\tau$-based $m$ in Figure \ref{fig:moment_vs_radius_tau} exhibit a slight nonlinearity, it is arguably appropriate to assign a linear fit $m(r) = C_1 + C_2 \cdot r$ linear fit, where $C_1 = (1.290 \pm 0.129) \cdot 10^{-16}~ A\cdot m^2$ and $C_2 = - (0.706 \pm 0.086) \cdot 10^{-10}~ A\cdot m^3$, with $m(r_c)=0$ for $r_c \sim 0.55 ~\mu m$.

The results obtained using both methods are compared side by side in Figures \ref{fig:moment_histogram_comparison_L_vs_tau} and \ref{fig:moment_vs_radius_comparison_L_vs_tau}. The $\tau$-based measurement method yields systematically lower $m$ values, and the probability distribution of $m$ is skewed towards larger values, more so than in the case of estimates based on $L$. In the latter, $m$ is characterized by both a larger range and larger values. However, we would like to reiterate that reducing the overall $m$ statistics is not representative of the diversity of the MTB properties in the population, which is better characterized by the scaling of $m(r)$. In Figure \ref{fig:moment_vs_radius_comparison_L_vs_tau} the slopes for $m(r)$ between the $L$- and $\tau$-based $m$ estimation cases differ very significantly, even thought the zero-moment effective radii $r_c$ are not that different. Since our method captures the spatial configuration of the U-turns, there is a much more diverse set of $m$ values at every $r$. However, because of the banding artifacts in the $\tau$-based data and, arguably, the more precise/physical treatment of trajectories and U-turns with \eqref{eq:uturn-shape-equation-final} and \eqref{eq:moment-from-uturn-params}, we posit that our method yields results more representative of the observed MTB population. Note that both the $m(r)$ datasets and the linear fits indicate a similar $r_c$ value, meaning that no magnetic cells smaller than $r_c$ were found in the population (that is, MTB with $r<r_c$ did not exhibit a significant U-turn motion).

Since the $L$-based $m$ measurement method contains the characteristic U-turn velocity $v$ as a parameter, we tested the same dataset as shown in Figures \ref{fig:m_moment_histogram} and \ref{fig:moment_vs_radius} using three different approaches to determine cell velocity -- the effects of different $v$ estimates on the $m$ probability statistics are shown in Figure \ref{fig:moment_diff_velocity_estimates}. Here, $m$ is calculated using the individual U-turn $v$ computed using Algorithm \ref{alg:mtb-moment-extraction}, the mean $v$ for all U-turns, and a simpler estimate based on the maximum probability velocity for the entire MTB population. Individual and mean U-turn $v$ estimates yield similar $m$ statistics, with the mean U-turn $v$ producing slightly less probable peaks at  $m \sim 0.6~ \cdot 10^{-16}~ A \cdot m^2$ and  $m \sim 0.9~ \cdot 10^{-16}~ A \cdot m^2$. However, using the most probable population velocity, the $m$ statistics are significantly different: $m$ distribution is more dispersed and with a significantly shifted mean and most probable values, indicating that one cannot neglect the $v$ calculations for individual U-turns in favor of a coarser, more readily available estimate.

Finally, it is also important to compare our results with the data available in the literature for MSR-1 MTB. Several approaches have been used to measure and calculate $m$. Tracking and magnetic tweezer-based methods reported in \cite{Pichel_2018} and \cite{zahn_measurement_2017}, respectively, yield lower $m$ values than those calculated from TEM images \cite{zahn_measurement_2017}. A study by Nadkarni \textit{et al.} \cite{nadkarni_comparison_2013} indicates that for AMB-1, also a spirillum and arguably similar to MSR-1, the $m$ values calculated using the $\tau$-based U-turn method were close to those obtained by magnetosome volume measurement. However, it is known that the growth conditions of the MTB population (e.g. oxygen concentration \cite{heyen_growth_2003} and iron content of the medium can influence the characteristics of the cell population, including cell size and magnetosome formation \cite{fast_growing_magnetosomes}. Therefore, it is reasonable to expect a significant variation of $m$ values between different cultures and experiments.

Table \ref{tab:m_comparison_with_literature} summarizes how our results compare with those reported for MSR-1. The first thing to note is that our mean values for $m$ using both $L$- and $\tau$-based methods are within the expected order of magnitude, $m \sim 10^{-16}~ A\cdot m^2 $. However, they are also significantly lower than the rest of the presented $m$ values. There could be a number of reasons for this, including inherently lower $m$ for our MSR-1 population due to the way magnetosomes developed under the conditions defined by our culture preparation. While the velocimetry results (Figures \ref{fig:velocity_histogram} and \ref{fig:velocity_radius_histogram_hexplot}) for the population indicate that the observed MTB move with velocity close to what is reported by Pitchel \textit{et al.} \cite{Pichel_2018}, there were relatively many bacteria in the lower end of the velocity range. Importantly, Pitchel \textit{et al.} performed measurements indicating that the spiral shape of the body of an
MTB has a $64 \pm 5 \%$ higher (rotational) drag $\alpha$ than a spheroid with equivalent length
and diameter \cite{Pichel_2018}. However, even factoring in this observation and noting that $m \sim \alpha$ in \eqref{eq:moment-from-uturn-params}, and that 

\begin{equation}
    \frac{m}{\ln{m}} \sim \frac{\alpha}{\tau} 
\end{equation}
in \eqref{eq:magnetic-moment-tau}, our expected $m$ values from $L$-based and $\tau$-based calculations still do not exceed $\sim 1.37 \cdot 10^{-16}~ A \cdot m^2$ and $\sim 0.58 \cdot 10^{-16}~ A \cdot m^2$, respectively (also recorded in Table \ref{tab:m_comparison_with_literature}). Therefore, this correction does not explain all the difference between our $L$- and $\tau$-based results, and does not account for a significant part of the difference between our data and the reported MSR-1 results.

This means that spontaneous loss of magnetic properties of our MSR-1 population during the growth process must be considered \cite{le_nagard_growing_msr_1, magn_phenotype_loss}, especially since the U-turn decomposition and $m$ calculation procedure (Algorithm \ref{alg:mtb-moment-extraction}) filtered out very many tracks corresponding to MTB that do not respond to MF. These include MTB that are stuck, moving in patterns uncorrelated to MF direction or MF switching, traveling straight through the FOV, exhibiting only minor large-wavelength trajectory perturbations during MF switching, etc. This is reflected in the fact that the $\sim 18~\text{K}$ trajectories retrieved from the $1~\text{K}$ images yielded $N=304$ U-turns that Algorithm \ref{alg:mtb-moment-extraction} determined as viable based on the criteria explained in Section \ref{sec:determining-L} and summarized in Appendix \ref{appendix:moment-calculation-algorithm-parameters}. We have therefore analyzed another population of MSR-1 -- $1~\text{K}$ images were acquired and filtered to $\sim 3.3~\text{K}$ MTB trajectories, resulting in $N=45$  $m$ values with the mean value $m = 0.49 \pm 0.4 \cdot 10^{-16}~ A \cdot m^2$ ($L$-based calculation \eqref{eq:moment-from-uturn-params}), corresponding to $r = 0.70 \pm 0.02 ~\mu m$ (our MTB length is $l = 3.14 \pm 0.34$ and width is $d = 0.24 \pm 0.06$, as seen in Table \ref{tab:m_comparison_with_literature}). Here, the $r$ statistics match what we show in Table \ref{tab:m_comparison_with_literature} and the $m$ values are of the same order of magnitude, indicating consistency despite the perhaps lower viability of the control batch. However, we have chosen not to combine the data with the population data in Table \ref{tab:m_comparison_with_literature} since the velocity statistics were significantly different, therefore only using the smaller $m$ value batch as a reference. Note that the number of $m$ measurement instances we have acquired exceeds the dataset size from any other source for MSR-1, which allowed us to assess the $m(r)$ dependency, so data sparsity is not the issue here. In addition, converting the MTB dimensions specified in Table \ref{tab:m_comparison_with_literature} from the other sources on MSR-1 indicates that the observed MTB were of similar size. Given all of the above, it is reasonable to assume that the differences in the values of $m$ are probably mainly due to the different growth conditions of MSR-1 prior to the experiments.

\begin{table}
    \begin{center}
    \begin{tblr}{
      cells={valign=m,halign=c},
      row{1}={bg=custompink,font=\bfseries},
      colspec={|c|c|c|c|}
    }
         \hline
         $\boldsymbol{m, 10^{-16}~ A \cdot m^2}$&  Method& Source& MTB size, $\boldsymbol{\mu m}$\\
         \hline
         $0.23 \pm 0.01$; $N=304$ &  U-turn, $\tau$-based equation \eqref{eq:magnetic-moment-tau} & This paper & $l = 3.14 \pm 0.34$\\ & & & $d = 0.24 \pm 0.06$ \\
         \hline
         $0.78 \pm 0.03$; $N=304$ &  U-turn, $L$-based equation \eqref{eq:uturn-shape-equation-final} & This paper & $l = 3.14 \pm 0.34$\\ & & & $d = 0.24 \pm 0.06$ \\
         \hline
         \SetCell[r=3]{m} $0.58\pm 0.03$; $N=304$ &  U-turn, $\tau$-based equation \eqref{eq:magnetic-moment-tau} & \SetCell[r=3]{m} This paper & \SetCell[r=2]{m} $l = 3.14 \pm 0.34$ \\  & \SetCell[r=2]{m} Drag correction from \cite{Pichel_2018} & & \\ & & & $d = 0.24 \pm 0.06$ \\
         \hline
         \SetCell[r=3]{m} $1.37 \pm 0.05$; $N=304$ &  U-turn, $L$-based equation \eqref{eq:uturn-shape-equation-final} & \SetCell[r=3]{m} This paper & \SetCell[r=2]{m} $l = 3.14 \pm 0.34$ \\  & \SetCell[r=2]{m} Drag correction from \cite{Pichel_2018} & & \\ & & & $d = 0.24 \pm 0.06$ \\
         \hline
         \SetCell[r=3]{m} $2.5 \pm 0.5$; $N=174$ &  U-turn diameter-based & \SetCell[r=3]{m} Pitchel \textit{et al.} \cite{Pichel_2018} & \SetCell[r=2]{m} $l=5.0 \pm 0.2$ \\  & \SetCell[r=2]{m} Semi-manual, no shape fitting & & \\ & & & $d=0.24 \pm 0.01$ \\
         \hline
         \SetCell[r=3]{m} $2.4 \pm 1.1$; $N=265$ &  Rotation in alternating MF & \SetCell[r=3]{m} Zahn \textit{et al.} \cite{zahn_measurement_2017} & \SetCell[r=2]{m} $l=3.7 \pm 0.7$ \\  & \SetCell[r=2]{m} Magnetic tweezers & & \\ & & & $d=0.42 \pm 0.03$ \\
         \hline
         \SetCell[r=3]{m} $7.7 \pm 3.4$; $N=86$ &  Translation in constant MF & \SetCell[r=3]{m} Zahn \textit{et al.} \cite{zahn_measurement_2017} & \SetCell[r=2]{m} $l=3.8 \pm 0.7$ \\  & \SetCell[r=2]{m} Magnetic tweezers & & \\ & & & $d=0.42 \pm 0.03$ \\
         \hline
         $9.9 \pm 2.6$; $N=50$&  Magnetosome volumes via TEM & Zahn \textit{et al.} \cite{zahn_measurement_2017}&N/A\\
         \hline
         $3.54 \pm 2.65$; $ N=102 $ &  Magnetosome volumes via TEM & Codutti \textit{et al.} \cite{codutti_interplay_2022} & \ N/A \\
         \hline
    \end{tblr}
    \end{center}
    \label{tab:m_comparison_with_literature}
    \caption{Comparison of different MSR-1 magnetic moment values.}
\end{table}

It is important to point out that our approach, which treats the MTB trajectory and U-turn shapes explicitly based on a physical model, is also fully automated start-to-finish: image processing, MTB tracking, eligible trajectory selection, U-turn decomposition, U-turn selection, and $m$ calculations. This serves to minimize or eliminate human error and bias. This is, for instance, in contrast to what is reported in \cite{Pichel_2018}, where a much less robust image MTB detection algorithm is presented that may not work for the number density of MTB in images as considered here, or in the presence of diffraction artifacts at MTB locations or other dynamic artifacts in the FOV. Moreover, nearest neighbor tracking is used \cite{Pichel_2018}, which is known to be very unstable at higher MTB number densities, and when multiple MTB are in proximity and/or their paths are spatiotemporally similar or are crossing. This reduced data acquisition efficiency (less viable MTB tracks are reconstructed) and track reconstruction errors are more frequent and significant. More advanced methods are required for general-purpose MTB tracking, and for the U-turn method case in particular -- for instance, approaches like shake-the-box \cite{shake-the-box-lagrangian-tracking} (robust, thoroughly validated, and very widely used in the hydrodynamics community -- can be used for long track reconstruction) and multiple hypothesis tracking (MHT), specifically its offline version MHT-X used here \cite{mht-x-og,birjukovs-particle-EXIF,birjukovs-particle-track-curvature-stats} with motion modeling based on weak conservation laws, and velocimetry-assisted motion prediction with boundary condition support. Other relevant methods are reviewed in \cite{lagrangian-tracking-fluid-mechanics-review}.

In addition, the eligible track selection process is reported as semi-manual, which precludes larger-scale data analysis, such as that reported here, and is inherently biased. The U-turn diameter $D$ used in $m$ calculations \cite{Pichel_2018} is determined manually -- trajectory fragments corresponding to the U-turn motion are segmented manually, and $D$ is determined as the U-turn endpoint distance, not the asymptotic branch distance as in \eqref{eq:uturn-shape-equation-final} or, simpler, by an ellipse or circle fit. Alternatively, this could have been done via curvature calculation and thresholding, as in Algorithm \ref{alg:mtb-moment-extraction} for U-turn $v$ calculation. Interestingly, despite a greater reported $m$ value and similar population velocity statistics, the U-turn diameters $D$ in \cite{Pichel_2018} at $\sim 1.85~mT$ are on average $\sim 32 \pm 4 ~\mu m$ (derived from data the reported in \cite{Pichel_2018}[Figures 24, 25]), whereas our statistics for $L$ at $2~mT$ indicate an average value of $\sim 7.7 \pm 3.7 ~ \mu m$. This seems to be more in line with U-turn numerical simulations performed in \cite{Pichel_2018}, where the U-turn width at $2~mT$ is $\sim 11~\mu m$. However, the properties of MTB used for these simulations are specified only as "realistic" \cite{Pichel_2018}, so direct comparison is difficult. We would like to stress again that comparing our $L$ values against $D$ can only be done very roughly, since the calculation procedures are vastly different, and $D$ is not strictly defined.

We have also noticed that MTB average (characteristic) velocity $v$ for $m$ calculations in \cite{Pichel_2018} is computed for the entire U-turn (i.e., turning arcs and branches), whereas our $m$ values are computed from $v$ sampled from the turning arcs (Figure \ref{fig:u-turn-decomposition}d). This might be important, as we have shown in Figure \ref{fig:moment_diff_velocity_estimates}. In our case, the mean U-turn velocity is $\sim 20\%$ smaller than the mean population velocity. If this is also true for the population considered in \cite{Pichel_2018}, then, with everything else unchanged, $m \sim (2.0 \pm 0.4) \cdot 10^{-16}~ A\cdot m^2$. This is still greater than our drag-corrected value $m = (1.37 \pm 0.05) \cdot 10^{-16}~ A\cdot m^2$, but given the error margins, the results are arguably close.

Among the $m$ measurements listed in Table \ref{tab:m_comparison_with_literature} for MSR-1, the U-turn method is only used in \cite{Pichel_2018}. It is then also important to consider other $m$ measurement cases that are based on U-turn methods, although they are not relevant to MSR-1. In simpler cases, methods deriving U-turns from trajectories via the extrema in the U-turn trajectory coordinate derivatives \cite{alvaros_u_turn_cocci,nadkarni_comparison_2013,sales_u-turn_2020,chevrier_collective_2023} may be sufficient (although errors and lower precision can be expected). In cases with longer and more complex trajectories, as well as with higher MTB number density in images, these methods should not be considered reliable and precise.

Even though the U-turn method is not used in \cite{codutti_interplay_2022}, the data presented in \cite{codutti_interplay_2022}[Figure 5b] and in the repository linked to the paper can still be used to estimate $m$. As shown in Table \ref{tab:m_comparison_with_literature}, the $m$ value due to TEM measurements is $(3.54 \pm 2.65) \cdot 10^{-16}~ A\cdot m^2$, and this value is used as a reference for calculations and data analysis in \cite{codutti_interplay_2022}. At the same time, using the available data, we recovered the MSR-1 MTB trajectory turning (not exactly the U-turns as defined here or in \cite{Pichel_2018}) radii within circular traps, and the respective MTB velocity values. Since the MTB length is not provided explicitly in \cite{codutti_interplay_2022}, but rather a body length scale is specified $l \sim 3~\mu m$, we can only provide a very rough estimate via \eqref{eq:moment-from-uturn-params}, assuming the same mean MTB size as in our case, and taking the $0.5~mT$ MF as in \cite{codutti_interplay_2022}. With this, we estimate $m \sim (2.08 \pm 0.07) \cdot 10^{-16}~ A\cdot m^2$ with a spherical drag model, and $m \sim (3.44 \pm 0.12) \cdot 10^{-16}~ A\cdot m^2$ if the correction from \cite{Pichel_2018} is accounted for, although the error margins cannot be rigorously estimated (e.g., velocity and turning radii uncertainties not specified in \cite{codutti_interplay_2022}), and are likely much wider. Note that for MTB with different dimensions, the estimate will differ significantly \cite{swimming-organism-data-bank}. This and the strongly confined motion (turning radius affected by trap boundary conditions) makes the comparisons with our results and the data from \cite{Pichel_2018} problematic, but validates the order of magnitude for the measured $m$ values.

Arguably, given the current state of literature on MTB $m$ measurements (relatively few measurements per research paper \cite{zahn_measurement_2017, Pichel_2018, alvaros_u_turn_cocci,chevrier_collective_2023,codutti_interplay_2022} and little to no discussion of $m(r)$ scaling \cite{alvaros_u_turn_cocci}), and the diversity of MTB types and their properties (which depend on the growth conditions) there is a need for an extensive measurement campaign where many different cultures of various MTB are analyzed systematically and automatically, to obtain likely thousands of magnetic moment measurement instances. The U-turn method is promising in this regard, since it can be used for simultaneous $m$ measurements for many MTB, and long image sequences can be acquired for processing. The methodology we have presented here, unlike the U-turn time-based calculations, works with U-turn shapes directly, and thus more precisely, and is readily scalable to measuring $m$ from many thousands of images with little to no parameter adjustment. Most of the underlying code is efficiently parallelized and can benefit from high-performance computing hardware.

\section{Conclusions \& outlook}
\label{sec:conclusions}

Using the proposed methodology, we performed optical microscopy experiments with high number density motion of MSR-1 MTB in alternating MF, and determined the magnetic moment for 304 MTB. Our results show that computing $m$ from time-based methods underestimates the magnitude of $m$ compared to our method based on the distance between trajectory branches $L$. Our results are within the order of magnitude of the data reported in the literature (closer after adjusting for a helical MTB drag model), and the differences are possibly due to spontaneous mutations that can cause culture to lose magnetic properties. We have also shown the scaling of MTB magnetic moment with MTB size, which is roughly linear for our data -- to our knowledge, such data is currently missing for widely used MSR-1 MTB.

To achieve the above, we have derived a theoretical shape function for the U-turn motion of magnetotactic microorganisms, including MTB, in alternating MF, and used it to improve the precision of the U-turn method used to determine the magnetic moment for magnetotactic microorganisms. We have also presented an alternative expression for computing the magnetic moment, which is based on the asymptotic width of the shape function, the MTB rotational drag coefficient, alternating MF strength, and the characteristic U-turn velocity. This method, unlike the U-turn time-based ($\tau$) moment calculation, determines the magnetic moment explicitly from the U-turn geometry and uses the entire U-turn length, without the need to compute $\tau$.

We have coupled the new shape function with a fully automated method for MTB trajectory decomposition into U-turns, which also determines the characteristic U-turn velocity and then computes the magnetic moment by fitting the theoretical U-turn shape function to the experimentally observed U-turns.

We have reported significant differences in the magnetic moment statistics and scaling with MTB size between the established $\tau$-based moment calculation and the shape function-based method proposed here, and highlighted two important limitations of the $\tau$-based method -- its much stronger dependence on higher image acquisition frame rate to sample the magnetic moment statistics, and its lower precision since it does not model the U-turn geometry. We have also shown that it is important to determine the characteristic U-turn velocity values explicitly for each U-turn, since the magnetic moment statistics based on these differ significantly from the calculation based on a cruder population-based estimate.

We argue that the U-turn method is very promising for simultaneous measurement of magnetic moment for many MTB from a sequence of images, given that the method is paired with sufficiently robust, precise, and automated methods for MTB detection, tracking, and U-turn analysis. We posit that our U-turn shape function-based magnetic moment calculation, combined with the proposed fully automated U-turn decomposition and analysis approach, will enable convenient property determination with many more data points than currently reported (magnetic moment distribution, magnetic moment scaling with MTB size, etc.) for populations of different magnetotactic microorganisms. The proposed methods are scalable to automatically analyze thousands of images and tens of thousands of trajectories, all the relevant code is open source, and the parameters are provided either in this paper or with the code.

Lastly, we propose to apply the methods developed in this paper to process a large dataset of images with MSR-1 motion in alternating MF to refine the current results. This should be continued by a systematical study how the utilized rotational drag model affects the magnetic moment statistics and scaling with MTB size by considering the spirillum drag model experimentally established by Pichel et al. \cite{Pichel_2018}, as well as the cylindrical \cite{popp_polarity_2014}, ellipsoidal \cite{leao_eliptic_drag_c} and the sphere chain \cite{mtb-rotational-drag-sphere-chain-model} MTB shape models.

\section*{Acknowledgments}

The authors are grateful to Dr. Paul Janmey from the University of Pennsylvania for his feedback and editing of the manuscript.

\subsection*{Data \& materials availability}

Experimental data is available on reasonable request -- please contact the corresponding author.

\subsection*{Code availability}

The image analysis, object tracking, and trajectory analysis code for $m$ calculations are available on \textit{GitHub}:

\begin{itemize}[noitemsep]
    \item MTB detection from images: \href{https://github.com/Mihails-Birjukovs/MTB_detection_tracking}{Mihails-Birjukovs/MTB\_detection\_tracking}
    \item MTB tracking: \href{https://github.com/Peteris-Zvejnieks/MHT-X/tree/particle_tracking_PIV_assisted}{Peteris-Zvejnieks/MHT-X}
    \item MTB magnetic moment retrieval from trajectories: \\ \href{https://github.com/Mihails-Birjukovs/MTB_magnetic_moment_from_U-turn_trajectories}{Mihails-Birjukovs/MTB\_magnetic\_moment\_from\_U-turn\_trajectories}
\end{itemize}

\subsection*{Authors' contributions}

Mara Smite (M.S.) prepared the MTB culture, set up and performed the experiments; M.S. and Mihails Birjukovs (M.B.) performed data analysis and interpretation; M.B. developed the image processing, trajectory/U-turn analysis, and magnetic moment calculation methods and code; Peteris Zvejnieks (P.Z.) contributed the trajectory reconstruction software and assisted with its application; Ivars Drikis (I.D.) developed the real-time magnetic field recording setup; Andrejs Cebers (A.C.) derived the theoretical U-turn shape function; Guntars Kitenbergs (G.K.) provided funding and resources, and contributed to result interpretation; M.S., M.B., A.C. and G.K. contributed to the first draft. All authors reviewed and edited the final version of the manuscript.

\printbibliography[title={References}]

\begin{appendices}

\section{MTB $m$ retrieval from trajectories}
\label{appendix:moment-calculation-algorithm-parameters}

\begin{algorithm2e}
    \nonl \textbf{\underline{Input:}} MBT trajectories reconstructed with MHT-X
    
    \nonl \underline{\textit{Trajectory filtering}}
    
    \pushline Threshold by temporal length $N_t$, $N_t < N_{t1}$ cases eliminated
    
    Threshold by oriented aspect ratio $\rchi$, $\rchi < \rchi_{c1}$ cases eliminated
    
    Median and mean filtering, 1 point kernel width 
    
    \nonl \popline \underline{\textit{U-turn decomposition}}
    
    \pushline Compute second Gaussian derivatives ($\sigma_{c}$) of coordinate time series
    
    Threshold extrema points by relative magnitude $M$, removing $M<M_{c}$ cases, after Gaussian smoothing ($\sigma_{p}$)

    Compute the U-turn curvature with a Gaussian kernel ($\sigma_{k1}$)

    Remove extrema points with $M<M_{k1}$ after Gaussian smoothing ($\sigma_{t1}$) -- define as \textit{turning points}

    Combine the coordinate derivative extrema points with the turning points, resolve duplicates and proximity conflicts -- define as \textit{partition points}

    Partition tracks into U-turns at the partition points, keep partitions containing the turning points

    Prune the U-turns by $N_\text{cut}$ points from either endpoint

    \nonl \popline \underline{\textit{U-turn filtering}}
    
    \pushline Eliminate instances with $N_t < N_{t2}$
    
    Eliminate the self-intersecting U-turns

    Eliminate instances with $\rchi < \rchi_{c2}$

    Threshold by oriented minimum area $S$, $S < S_c$ cases eliminated

    Refine the turning points via Steps 6 \& 7 with parameters $\sigma_{k2}$, $M_{k2}$ and $\sigma_{t2}$, performing refinement on U-turns upsampled by a factor $K_1$

    Eliminate U-turns with multiple turning points

    \nonl \popline \underline{\textit{Determining U-turn $L$ values}}

    \pushline Symmetrize U-turn branches by physical length 

    Translate the turning points to $(x,y)=(0,0)$

    Eliminate instances with $N_t < N_{t3}$

    Eliminate U-turns with longest/shortest branch temporal length ratio $> \rchi_b$

    Resample "secondary" turning points via Steps 6 \& 7 with parameters $\sigma_{k3}$, $M_{k3}$ and $\sigma_{t3}$ with the upsampling factor $K_2$, eliminate instances with multiple secondary points

    Eliminate U-turns with MF alignment factor $A<A_c$

    Run the $L$ optimization routine \eqref{eq:optimization-fitness-total}

    Keep U-turns and $L$ values with $g > g_c$ \eqref{eq:optimization-penalty-function} and $f + \beta g > f_c$ \eqref{eq:optimization-fitness-total}
    
    \nonl \popline \underline{\textit{Computing $m$ values}}
    
    \pushline Propagate MTB property data from tracks to eligible U-turns
    
    Compute $\alpha$ \eqref{eq:mtb-drag-shpherical}

    Compute the mean-filtered (1 point-wide kernel) velocity time series

    Compute Gaussian-smoothed ($\sigma_{k4}$) curvature for U-turns

    Compute mean velocity $v$ for U-turn points with relative curvature $> M_{k4}$

    Compute $m$ via \eqref{eq:moment-from-uturn-params} from $\alpha$, $v$, and $L$

    \nonl \popline \textbf{\underline{Output:}} 
    \begin{itemize}[noitemsep,topsep=0pt]
    \popline \popline \popline \item MTB magnetic moment $m$ values with uncertainties
        \item Corresponding filtered U-turns
    \end{itemize}
    
\caption{Determining MTB magnetic moment $m$ from trajectory data}
\label{alg:mtb-moment-extraction}
\end{algorithm2e}

Parameters for Algorithm \ref{alg:mtb-moment-extraction}:
\begin{enumerate}[noitemsep]
    \item $N_{t1} = 10$, $\rchi_{c1} = 10$
    \item $\sigma_{c} = 5$, $M_{c} = 0.85$, $\sigma_{p} = 5$
    \item $\sigma_{k1} = 0.05 \cdot N_t$, $M_{k1} = 0.95$, $\sigma_{t1} = 2.5$
    \item $N_\text{cut} = \max{\left( 1; \nint{n_\text{cut} \cdot N_t} \right) },~ n_\text{cut} = 7.5 \cdot 10^{-2} $  
    \item $N_{t2} = N_{t1}$, $\rchi_{c2} = \rchi_{c1}/2$
    \item $S_c = 3.5 \cdot (z_x \cdot z_y)^{1/2}$, where $z_k$ are image dimensions
    \item $(\sigma_{k2},M_{k2},\sigma_{t2}) = (0.85,1 \cdot 10^{-3},0.25) * (\sigma_{k1},M_{k1},\sigma_{t1})$, $K_1 = 10$
    \item $N_{t3} = N_{t1}$, $\rchi_b = 1.75$
    \item $(\sigma_{k3},M_{k3},\sigma_{t3}) = (0.25,1.75 \cdot 10^{-3},0.25) * (\sigma_{k1},M_{k1},\sigma_{t1})$, $K_2 = K_1$
    \item $A_c = 2$
    \item $\beta = 1$, $g_c = 1 \cdot 10^{-5}$, $f_c = 1 \cdot 10^{-2}$
    \item $\sigma_{k4} = 0.25 \cdot N_t$, $M_{k4} = 0.25$
\end{enumerate}

\end{appendices}

\end{document}